%% file: ms.tex
\documentclass[aps,showpacs,nofootinbib, onecolumn]{revtex4-1}
	
	\pdfoutput=1

	\usepackage{amsmath,amssymb,amsfonts,mathtools}
	\usepackage{graphicx}
	\usepackage{upgreek}
	\usepackage{xcolor}
	\usepackage{standalone}
	\usepackage{tikz}
	\usetikzlibrary{positioning, patterns, calc, shapes.misc, 3d, shapes.arrows}
	\usepackage{xstring}
	\usepackage{pgf}
    \usepackage{subcaption}

	\graphicspath{{./},{./images/}}

	\newcommand{\cref}[1]{Ref.~\cite{#1}}

	\newcommand{\bea}{\begin{eqnarray}}
	\newcommand{\eea}{\end{eqnarray}}

	\newcommand{\F}{\mathcal{F}}
	\newcommand{\Fid}{\mathcal{F}^\text{id}}
	\newcommand{\Fex}{\mathcal{F}^\text{ex}}
	\newcommand{\Fmf}{\mathcal{F}^\text{ex}_\text{mf}}
	\newcommand{\FAO}{\mathcal{F}_\text{AO}}
	\newcommand{\tFAO}{\tilde{\mathcal{F}}_\text{AO}}
	\newcommand{\Flg}{\mathcal{F}_\text{lg}}
	
	\newcommand{\Phil}{\Phi^\text{loc}}
    \newcommand{\Phinl}{\Phi^\text{nonloc}}
    
	\newcommand{\kt}{k_\text{B}T}
	\newcommand{\ktc}{k_\text{B}T_\text{c}}
    \newcommand{\xem}{x_\text{em}}
    \newcommand{\gem}{\gamma_\text{em}}

	\newcommand{\rhoc}{\rho_\text{c}}
	\newcommand{\rhocb}{\rho_\text{c,b}}
	\newcommand{\sums}{\sum_{\pmb{s}}}
	\newcommand{\Ve}{V_\text{ext}}
	\newcommand{\pzd}{\Phi^\text{0D}}
	\newcommand{\mupa}{\mu_{\text{p},\alpha}}
	\newcommand{\mup}{\mu_{\text{p}}}
	\newcommand{\mupcan}{\mu_{\text{pc}^n,\alpha}}
	\newcommand{\mucoex}{\mu_{\text{coex}}}
	\newcommand{\rhopra}{\rho^\text{res}_{\text{p},\alpha}}
	\newcommand{\rhopa}{\rho_{\text{p},\alpha}}
	\newcommand{\rhopr}{\rho^\text{res}_{\text{p}} }
	\newcommand{\rhopcan}{\rho_{\text{pc}^n,\alpha}  }
	\newcommand{\rhopca}{\rho_{\text{pc},\alpha}  }
	\newcommand{\rhopc}{\rho_{\text{pc}}  }
	\newcommand{\rhopcb}{\rho_{\text{pc,b}}  }
	\newcommand{\rhocoex}{\rho_{\text{coex}}}

	\newcommand{\Npa}{{N_{\text{p},\alpha}}}
	\newcommand{\Npcan}{N_{\text{pc}^n,\alpha}}

\include{newc_manuel}

\begin{document}

	\title{A density functional for the lattice gas from fundamental measure theory}
	\author{M.~Maeritz and M. Oettel}
	\email{Email address: martin.oettel@uni-tuebingen.de}
	\affiliation{Institut f\"ur Angewandte Physik, Eberhard Karls Universit\"at T\"ubingen, Auf der Morgenstelle 10, D-72076 T\"ubingen, Germany}

	\begin{abstract}
We construct a density functional for the lattice gas / Ising model on square and cubic lattices based on lattice 
fundamental measure theory. In order to treat the nearest-neighbor attractions between the lattice gas particles,
the model is mapped to a multicomponent model of hard particles with additional lattice polymers where effective attractions 
between particles arise from the depletion effect. The lattice polymers are further treated via the introduction
of polymer clusters (labelled by the numbers of polymer they contain) such that the model becomes a multicomponent model
of particles and polymer clusters with nonadditive hard interactions. The density functional for this nonadditive hard model is 
constructed with lattice fundamental measure theory. The resulting bulk phase diagram recovers the Bethe--Peierls
approximation and planar interface tensions show a considerable improvement compared to the standard mean--field functional
and are close to simulation results in three dimensions. We demonstrate the existence of planar interface solutions
at chemical potentials away from coexistence when the equimolar interface position is constrained to arbitrary
real values. 
	\end{abstract}

	\maketitle

\input{sec_1_introduction.tex}

\input{sec_2_construction_dft.tex}

\input{sec_3_results.tex}
    \input{sec_4_outlook.tex}

\begin{appendix}

\input{app_A_polymerclusters.tex}

\input{app_B_construction_excess.tex}

\input{app_C_surfacetension.tex}

\end{appendix}        

\bibliographystyle{PRE}
\bibliography{references.bib}

\end{document}

%% file: newc_manuel.tex
\newcommand\nUp[1]{
	\begin{tikzpicture}[scale=0.3, idx/.style={scale=0.5}, baseline=0pt, site/.style={circle, draw=black, fill=white, thin, minimum size=0.4cm, inner sep=0pt, scale=0.3}]
	\IfEqCase{#1}{
		{i}{\node[site] (G1) at (0,0.1)  {};}
		{12}{\node[site] (G1) at (0,0)  {};
			\node[site] (G2) at (1,0)  {};
			\draw[black, thin] (G1) -- (G2);}
		{13}{\node[site] (G1) at (0,0)  {};
			\node[site] (G3) at (0.5,0.866)  {};
			\draw[black, thin] (G1) -- (G3);}
		{23}{\node[site] (G2) at (1,0)  {};
			\node[site] (G3) at (0.5,0.866)  {};
			\draw[black, thin] (G2) -- (G3);}
		{123}{\node[site] (G1) at (0,0)  {};
			\node[site] (G2) at (1,0)  {};
			\node[site] (G3) at (0.5,0.866) {};
			\draw[black, thin] (G1) -- (G2) --(G3) -- (G1)-- cycle;}
	}
	\end{tikzpicture}}

\newcommand{\nUpEval}[1]{
	\begin{tikzpicture}[scale=0.3, idx/.style={scale=0.5}, baseline=0pt, site/.style={circle,  fill=black, minimum size=0.4cm, inner sep=0pt, thick, scale=0.3}]
	\IfEqCase{#1}{
		{i}{\node[site] (G1) at (0,0.1)  {};
			\node[idx]  [below right=-4pt and -0.05 of G1] {$i$};}
		{12}{\node[site] (G1) at (0,0)  {};
			\node[idx]  [below left=-4pt and -0.05 of G1] {$1$};
			\node[site] (G2) at (1,0)  {};
			\node[idx]  [below right=-4pt and -0.05 of G2] {$2$};
			\draw[black, thin] (G1) -- (G2);}
		{13}{\node[site] (G1) at (0,0)  {};
			\node[idx]  [below left=-4pt and -0.05 of G1] {$1$};
			\node[site] (G3) at (0.5,0.866)  {};
			\node[idx]  [above right=-4pt and -0.05 of G3] {$3$};
			\draw[black, thin] (G1) -- (G3);}
		{23}{\node[site] (G2) at (1,0)  {};
			\node[idx]  [below right=-4pt and -0.05 of G2] {$2$};
			\node[site] (G3) at (0.5,0.866)  {};
			\node[idx]  [above right=-4pt and -0.05 of G3] {$3$};
			\draw[black, thin] (G2) -- (G3);}
		{123}{\node[site] (G1) at (0,0)  {};
			\node[idx]  [below left=-4pt and -0.05 of G1] {$1$};
			\node[site] (G2) at (1,0)  {};
			\node[idx]  [below right=-4pt and -0.05 of G2] {$2$};
			\node[site] (G3) at (0.5,0.866) {};
			\node[idx]  [above right=-4pt and -0.05 of G3] {$3$};
			\draw[black, thin] (G1) -- (G2) --(G3) -- (G1)-- cycle;}
	}
	\end{tikzpicture}
	}

\newcommand{\nTwo}[2][\pmb{s}]{
	\begin{tikzpicture}[scale=0.07, site/.style={scale=0.5, label distance=-3pt}, baseline=0pt]
	\IfEqCase{#2}{
		{1}{\coordinate[label={[site]225:$#1$}] (origin) at (0,-1);
			\path[thin, draw=black, fill=blue, fill opacity=0.3, line join = round]
			(origin)-- ++(2,0) arc (180:0:1) -- ++(2,0) -- ++(0,2) arc (270:450:1)-- ++(0,2)-- cycle;
			\path[thin, draw=black, fill=green, fill opacity=0.3] (origin) rectangle +(6,6);
			\fill[color=black] (origin) circle (0.5);}
		{2}{\coordinate[label={[site]225:$#1$}] (origin) at (0,-1);
			\path[thin, draw=black, fill=blue, fill opacity=0.3, line join = round]
			(origin)-- ++(2,0) arc (180:0:1) -- ++(2,0) -- ++(0,2) arc (270:450:1)-- ++(0,2)-- cycle;
			\path[thin, draw=black, fill=green, fill opacity=0.3] ($(origin)+(6,0)$) rectangle +(6,6);
			\fill[color=black] (origin) circle (0.5);}
		{3}{\coordinate[label={[site]225:$#1$}] (origin) at (0,-1);
			\path[thin, draw=black, fill=red, fill opacity=0.3, line join = round]
			(origin) -- ++(6,6) -- ++(-2,0) arc (0:180:1) --++(-2,0) -- ++(0,-2)  arc (90:-90:1) -- cycle;
			\path[thin, draw=black, fill=green, fill opacity=0.3] (origin) rectangle +(6,6);
			\fill[color=black] (origin) circle (0.5);}
		{4}{\coordinate (origin) at (0,2);
			\path[thin, draw=black, fill=red, fill opacity=0.3, line join = round]
			(origin) -- ++(2,0) arc (180:0:1) --++(2,0) -- ++(-6,-6) -- ++(0,2) arc (270:450:1)-- cycle;
			\path[thin, draw=black, fill=green, fill opacity=0.3] (origin) rectangle +(6,6);
			\coordinate[label={[site]225:$#1$}] (s) at ($(origin)-(0,6)$);
			\fill[color=black] (s) circle (0.5);}
		{5}{\coordinate[label={[site]225:$#1$}] (origin) at (0,-1);
			\path[thin, draw=black, fill=blue, fill opacity=0.3, line join = round]
			(origin)-- ++(2,0) arc (180:0:1) -- ++(2,0) -- ++(0,2) arc (270:450:1)-- ++(0,2)-- cycle;
			\fill[color=black] (origin) circle (0.5);}
		{6}{\coordinate[label={[site]225:$#1$}] (origin) at (0,-1);
			\path[thin, draw=black, fill=red, fill opacity=0.3, line join = round]
			(origin) -- ++(6,6) -- ++(-2,0) arc (0:180:1) --++(-2,0) -- ++(0,-2)  arc (90:-90:1) -- cycle;
			\fill[color=black] (origin) circle (0.5);}
		{7}{\coordinate[label={[site]225:$#1$}] (origin) at (0,-1);
			\path[thin, draw=black, fill=green, fill opacity=0.3] (origin) rectangle +(6,6);
			\fill[color=black] (origin) circle (0.5);}
	}
	\end{tikzpicture}
	}

\newcommand{\nOne}[2][s]{
	\begin{tikzpicture}[scale=0.07, site/.style={scale=0.5, label distance=-3pt}, baseline=0pt]
	\IfEqCase{#2}{
		{1}{\coordinate[label={[site]225:$#1$}] (origin) at (0,-1);
			\path[thin, draw=black, fill=blue, fill opacity=0.3, line join = round]
			(origin)--++(6,0) -- ++(0,3)-- ++(3,3)--++(-6,0)--++(-3,-3)-- cycle;
			\path[thin, draw=black, fill=green, fill opacity=0.3] (origin) rectangle +(6,6);
			\fill[color=black] (origin) circle (0.5);}
		{2}{\coordinate[label={[site]225:$#1$}] (origin) at (0,-1);
			\path[thin, draw=black, fill=blue, fill opacity=0.3, line join = round]
			(origin)--++(6,0) -- ++(0,3)-- ++(3,3)--++(-6,0)--++(-3,-3)-- cycle;
			\path[thin, draw=black, fill=green, fill opacity=0.3] ($(origin)+(6,0)$) rectangle +(6,6);
			\fill[color=black] (origin) circle (0.5);
			\fill[color=black] ($(origin)+(6,0)$) circle (0.5);}
		{3}{\coordinate[label={[site]225:$#1$}] (origin) at (0,-1);
			\path[thin, draw=black, fill=blue, fill opacity=0.3, line join = round]
			(origin)--++(6,0) -- ++(0,3)-- ++(3,3)--++(-6,0)--++(-3,-3)-- cycle;
			\fill[color=black] (origin) circle (0.5);}
		{4}{\coordinate[label={[site]225:$#1$}] (origin) at (0,-1);
			\path[thin, draw=black, fill=green, fill opacity=0.3] (origin) rectangle +(6,6);
			\fill[color=black] (origin) circle (0.5);}
	}
	\end{tikzpicture}
}

\newcommand{\cavTwo}[1]{
	\begin{tikzpicture}[scale=0.07, site/.style={scale=0.5, label distance=-3pt}, baseline=0pt]
		\tikzset{
		cavity/.style={preaction={fill=gray, opacity=0.3}, pattern = north west lines}
		}
		\IfEqCase{#1}{
			{1}{\coordinate (origin) at (0,-1);
				\draw [very thin, black!50, step=6,xshift=0cm, yshift=-1cm] (-1,-1) grid +(8,8);
				\path[cavity] ($(origin)-(1,1)$) rectangle +(8,8)
				(origin)-- ++(6,0) -- ++(0,2)  arc (270:450:1)-- ++(0,2)--++(-6,0)-- cycle;
				\path[thin, draw=black, line join = round]
				(origin)-- ++(6,0) -- ++(0,2)  arc (270:450:1)-- ++(0,2)--++(-6,0)-- cycle;
				\coordinate[label={[site]225:$\pmb{r}$}] (s1) at (origin);}
			{2}{\coordinate (origin) at (0,-1);
				\draw [very thin, black!50, step=6,xshift=0cm, yshift=-1cm] (-1,-1) grid +(14,8);
				\path[cavity] ($(origin)-(1,1)$) rectangle +(14,8)
				(origin)-- ++(12,0) -- ++(0,6)--++(-6,0)-- cycle;
				\path[thin, draw=black, line join = round]
				(origin)-- ++(12,0) -- ++(0,6)--++(-6,0)-- cycle;
				\coordinate[label={[site]225:$\pmb{r}$}] (s1) at (origin);
				\coordinate[label={[site]225:$\pmb{s}$}] (s2) at ($(origin)+(6,0)$);
				}
			{3}{\coordinate (origin) at (0,-1);
				\draw [very thin, black!50, step=6,xshift=0cm, yshift=-1cm] (-1,-1) grid +(8,8);
				\path[cavity] ($(origin)-(1,1)$) rectangle +(8,8)
				(origin)-- ++(6,0) -- ++(0,6)-- ++(-2,0)  arc (0:180:1)-- ++(-2,0)--++(0,-6)-- cycle;
				\path[thin, draw=black, line join=round]
				(origin)-- ++(6,0) -- ++(0,6)-- ++(-2,0)  arc (0:180:1)-- ++(-2,0)--++(0,-6)-- cycle;
				\coordinate[label={[site]225:$\pmb{r}$}] (s1) at (origin);
				}
			{4}{\coordinate (origin) at (0,-4);
				\draw [very thin, black!50, step=6,xshift=0cm, yshift=-4cm] (-1,-1) grid +(8,14);
				\path[cavity] ($(origin)-(1,1)$) rectangle +(8,14)
				(origin)-- ++(6,6) -- ++(0,6)-- ++(-6,0)--cycle;
				\path[thin, draw=black, line join = round]
				(origin)-- ++(6,6) -- ++(0,6)-- ++(-6,0)--cycle;
				\coordinate[label={[site]225:$\pmb{r}$}] (s1) at (origin);
				\coordinate[label={[site]225:$\pmb{s}$}] (s2) at ($(origin)+(0,6)$);
				}
			}
	\end{tikzpicture}
	}

\newcommand{\cavOne}[1]{
	\begin{tikzpicture}[scale=0.07, site/.style={scale=0.5, label distance=-3pt}, baseline=0pt]
	\tikzset{
		cross/.style={cross out, draw=black, minimum size=2*(4pt-\pgflinewidth), inner sep=0pt, outer sep=0pt, scale=0.5}
	}
	\IfEqCase{#1}{
		{1}{\coordinate (origin) at (0,-1);
			\path[preaction={fill=gray, opacity=0.3}, pattern = north west lines] ($(origin)-(1,1)$) rectangle +(11,8);
			\path[thin, draw=black, fill=white, fill opacity=1, line join = round]
			(origin)--++(6,0) -- ++(0,3)-- ++(3,3)--++(-9,0)-- cycle;
			\coordinate[label={[site]225:$r$}] (s1) at (origin);
			\draw (origin) node[cross]{};}
		{2}{\coordinate (origin) at (0,-1);
			\path[preaction={fill=gray, opacity=0.3}, pattern = north west lines] ($(origin)-(1,1)$) rectangle +(14,8);
			\path[thin, draw=black, fill=white, fill opacity=1, line join = round]
			(origin)--++(12,0)--++(0,6)--++(-9,0)--++(-3,-3)--++(0,-3);
			\coordinate[label={[site]225:$r$}] (s1) at (origin);
			\coordinate[label={[site]225:$s$}] (s2) at ($(origin)+(6,0)$);
			\draw (origin) node[cross]{};
			\draw (s2) node[cross]{};}
	}
	\end{tikzpicture}
}

%% file: sec_1_introduction.tex
	\section{Introduction}

	The lattice gas model on a simple square or cubic lattice is defined 
	in terms of hard particles with mutual exclusion on the same lattice site and which have nearest--neighbor attractions of strength
	$-\epsilon$.
	It is equivalent to the Ising model on the respective lattices \cite{Huang1987}.
	The Ising model is one of the most intensely investigated models, with several exact results
	available in two dimensions (2D) and approximate, yet very precise results in three dimensions
	(3D) which have been obtained in many circumstances by simulation methods (see e.g. Ref.~\cite{Landau2018} for
	a recent, very precise estimate of critical properties in the 3D Ising model). 

	In equilibrium, the general inhomogeneous problem is defined by an arbitrary space--dependent 
	one--body external potential which 
	is in one--to--one correspondence with a resulting equilibrium density distribution (or profile).
	This entails the existence of a density functional for the grand potential whose
	minimization gives the equilibrium density distribution. Unfortunately, neither for the 2D or the
	3D lattice gas we know the explicit form for this functional.
	A different approach to solve the inhomogeneous problem is classical density matrix renormalization group theory \cite{Nishino1999}, utilizing
	transfer matrix methods, but currently solutions can be obtained for certain simple geometries (e.g. planar walls
	\cite{Drzewinski2009} or square ones \cite{Drzewinski2012}).

	In this article, we derive an approximate density functional 
	using a representation of the lattice gas in terms of
	a multi--species mixture of hard--core particles (colloids) with ideal gas lattice polymers (this is 
	a lattice version of the well--known continuum Asakura--Oosawa model \cite{Asakura1954}). The attraction between
	the  colloid particles effectively arises through the depletion effect \cite{Asakura1958}.
	The ideal gas lattice polymers are further mapped exactly to a multispecies system of polymer clusters
	(each polymer cluster contains a definite number of polymers and thus different polymer clusters exclude
	each other). This idea has been introduced in Ref.~\cite{Cuesta2005}. 
	Finally, the multicomponent model of colloids and polymer clusters is treated using
	lattice fundamental measure theory (FMT) \cite{Lafuente2002,Lafuente2004}.

	It has been demonstrated in Ref.~\cite{Cuesta2005} that this procedure gives an exact functional for the lattice gas in 1D. 
	However, in 2D and 3D it is not exact for reasons that will be detailed below. Nevertheless, it constitutes
	a major improvement beyond standard mean--field theory for the lattice gas which is widely used
	in conceptual modelling \cite{Archer2014} and which entails the famous Bragg--Williams approximation for the bulk
	phase diagram \cite{Huang1987}. Our functional will deliver an analytic form for the free energy and the equation of state in
	the Bethe--Peierls (or quasichemical) approximation. The associated bulk phase diagram 
	is already much closer to exact (2D) or quasiexact results (3D). With this functional, we study
	the free liquid--vapor interface and its tension, and we find the following effect presumably peculiar to lattice models. 
	A system with an equilibrium interface is only translationally invariant in discrete steps. However, if the 
	location of the interface is fixed at arbitrary positions away from these equilibrium positions, it connects
	liquid and vapor states off--coexistence, thus the interface sustains a pressure difference between the bulk phases.
	Indirect evidence for this has been seen in simulations before, without having been mentioned explicitly \cite{Troester2005,Troester2005a}.

	The paper is structured as follows. In Sec.~\ref{sec:construction} we present the construction of the
	density functional for the lattice gas which proceeds via showing the equivalence to a suitable lattice 
	model, the reformulation of the polymer ideal gas in terms of hard--core polymer clusters and finally
	the lattice FMT treatment of the mixture of lattice gas particle and polymer clusters. In 
	Sec.~\ref{sec:results} we show the analytical equation of state for the model and determine the
	phase diagram. Planar interface tensions between the gas and liquid phase are determined and we discuss
	the solutions for constrained interfaces off--coexistence. Finally, Sec.~\ref{sec:outlook} gives some 
	conclusions and an outlook.

%% file: sec_2_construction_dft.tex
	\section{Construction of the density functional}
    \label{sec:construction}
     
	\subsection{Mean--field functional}

	We define an ensemble--averaged density $\rhoc(\pmb{s})$ for lattice gas particles (where the index ``c'' stands for colloid)
	on discrete lattice sites $\pmb{s}$ of a square lattice (2D) or a simple cubic lattice (3D). In density functional theory,
	one defines a functional for the grand potential \cite{Evans1979}
	\bea
	  \Omega[\rhoc] = \Fid[\rhoc] + \Fex[\rhoc] - \sums (\mu - \Ve(\pmb{s})) \rhoc(\pmb{s})
	\eea
	which is split into an ideal gas free energy functional $\Fid$, an excess free energy functional $\Fex$ and a remaining part
	containing the chemical potential $\mu$ and the contribution of a one--body external potential $\Ve(\pmb{s})$. The ideal
	gas free energy functional is given by
	\bea
	  \beta \Fid[\rhoc] = \sums \rhoc(\pmb{s}) ( \ln \rhoc(\pmb{s}) -1)
	\eea
	(where $\beta= 1/(\kt)$ is the inverse temperature) and the excess functional is generally unknown. For hard core lattice particles  
	(i.e. $\epsilon=0$), however, the exact functional is known and given by
	\bea
 	  \beta \Fex_\text{hc}[\rhoc] = \sums \left( \rhoc(\pmb{s}) + (1-\rhoc(\pmb{s})) \ln(1-\rhoc(\pmb{s})) \right) =: {\sums} \pzd(\rhoc(\pmb{s}) ) \;.
	 \label{eq:fexhc}
	\eea
	Here we have introduced the free energy of a zero--dimensional (0D) cavity $\pzd(\eta)$ which can hold at most one particle and whose
	average occupation is given by $\eta \in [0,1]$. It plays a prominent role in the construction of lattice FMT \cite{Lafuente2002,Lafuente2004}.

	For $\epsilon >0$, the excess functional can be approximated by a standard mean--field treatment
	\bea
	  \Fmf[\rhoc] = \Fex_\text{hc}[\rhoc] - \frac{\epsilon}{2} \sums \sum_{\pmb{s}' \in \text{n.n.}(\pmb{s})} \rho(\pmb{s}) \rho(\pmb{s}') \;,  
	 \label{eq:fmf}
	\eea
	where the summation over lattice points $\pmb{s}'$ is restricted to nearest neighbors of $\pmb{s}$ ($\text{n.n.}(\pmb{s})$).

	The total free energy can be evaluated for a homogeneous bulk density and the phase diagram can be constructed: this gives the
	Bragg--Williams approximation \cite{Bragg1934,Huang1987}. The critical temperature is given by 
	\bea
	  \ktc = \frac{z}{4}\; \epsilon \;,
	\eea 
	where $z$ is the number of nearest neighbors of a lattice site ($z=4$ in 2D and $z=6$ in 3D). Thus, in 2D
	we obtain $\ktc= \epsilon$ which can be compared with the exact Onsager value of $\ktc = \epsilon/ (2 \ln(1+ \sqrt{2}) ) \approx 0.57 \epsilon$ \cite{Onsager1944}.
	In 3D we have from mean field $\ktc= 1.5 \epsilon$ which can be compared with very precise simulation estimates, giving $\ktc \approx 1.12 \epsilon$ \cite{Landau2018}. 

	\subsection{Equivalence to a lattice AO model}

	\subsubsection{Lattice gas functional as an effective AO functional}
	\label{sec:effAO}

\begin{figure}
	  \centerline{\includegraphics[width=8cm]{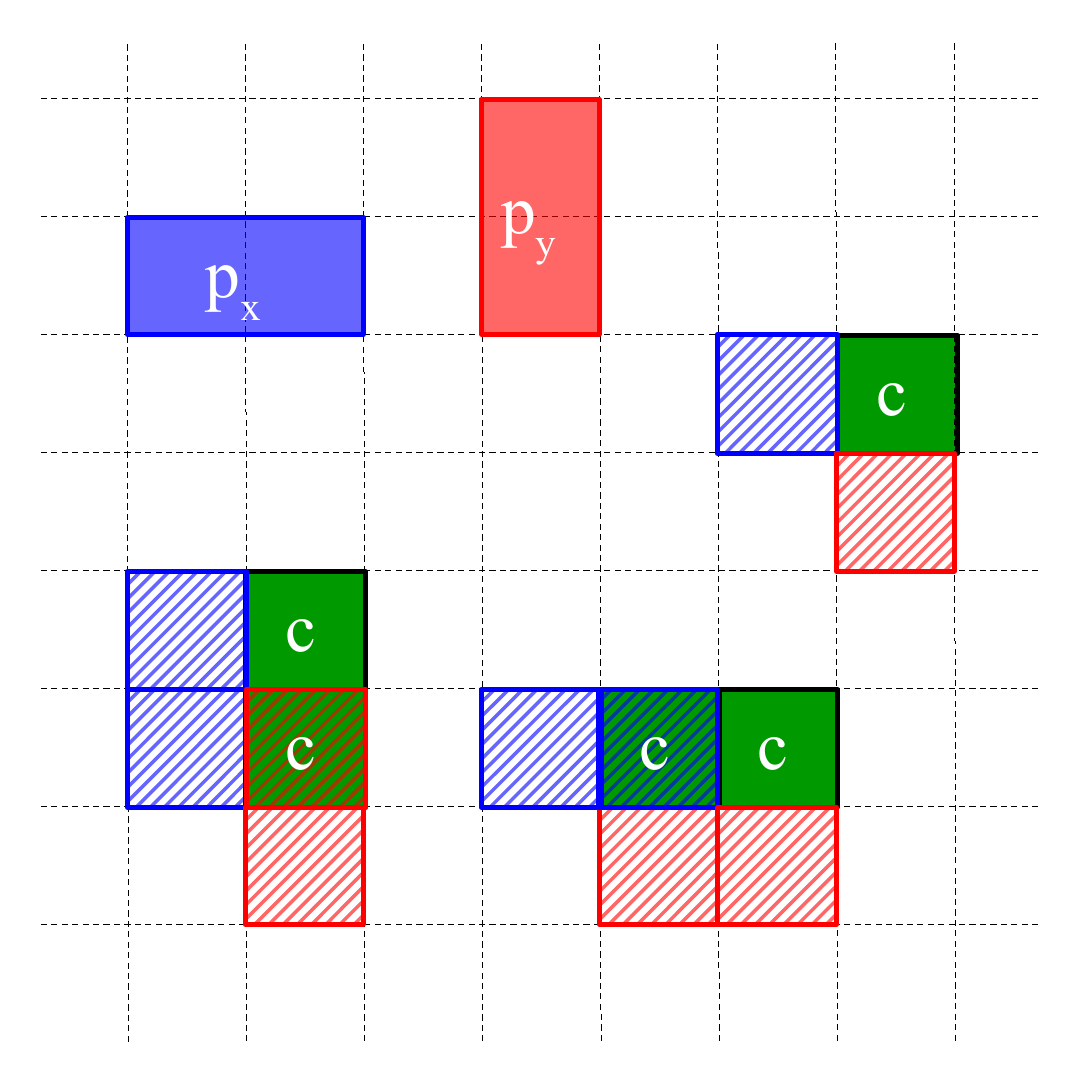}}
	  \caption{The definition of the AO lattice gas in 2D, see text for more details.}
	  \label{fig:aodef}
	\end{figure}

	In the following, we map the lattice gas to a lattice AO model with one colloid species and two polymer species (in 2D) and
	three polymer species (in 3D). For a visualization in 2D, see Fig.~\ref{fig:aodef}. The derivation in this subsection closely follows
	Ref.~\cite{Mortazavifar2017}.
	The colloid particles are hard core particles occupying one lattice site. The polymer particles are ideal gas particles but they interact hard
	with the colloid particles, equivalent to a rod occupying two neighboring lattice sites. Such a rod can be oriented in each Cartesian direction, therefore
	we have two species in 2D and three species in 3D. The lattice coordinate of a polymer particle is given by the one of the two covered lattice sites
	which has minimal coordinates (in 2D that would be the lower left lattice site). Therefore, a particular polymer cannot occupy the lattice site of
	a colloid at site $\pmb{s}$ and one neighboring lattice site (hatched in Fig.~\ref{fig:aodef}), these two sites define an exclusion volume 
	$V_\text{excl}(\pmb{s})$ for the polymer.

	The polymers couple to the colloids via a grand--canonical reservoir with a chemical potential $\mupa$ (where $\alpha=\{x,y\}$ in 2D and
	$\alpha=\{x,y,z\}$ in 3D). We also define reservoir polymer densities by $\rhopra=\exp(\beta\mupa)$. 

	When two colloids occupy neighboring lattice sites in the lattice direction $\alpha$, 
	their exclusion volumes overlap (with an overlap of exactly one lattice site),
	and this gives rise to a two--body depletion potential
	   $\beta V^\text{dep}_\alpha = - \rhopra \;.$
	For non--overlapping colloids, there are no triple and higher order overlaps possible between their exclusion volumes. Therefore
	$V^\text{dep}_\alpha$ is the only effective potential between colloids when polymers are integrated out. If we choose
	$\rhopr = \rhopra$ (i.e., equal chemical potential for all polymer species), the effective potential is isotropic in all
	lattice directions, and we recover the lattice gas upon identifying $\rhopr=\beta \epsilon$. 

	The mixture of colloids and polymers is described by a free energy functional $\FAO[\rhoc, \{\rhopa\}]$ which is split into an
	ideal gas and an excess part as before:
	\bea
	 \label{eq:fao}
	  \FAO[\rhoc, \{\rhopa\}] &=& \FAO^\text{id}[\rhoc, \{\rhopa\}] + \FAO^\text{ex}[\rhoc, \{\rhopa\}] \\
	  \beta \FAO^\text{id}[\rhoc, \{\rhopa\}] &=&  \sums \left(  \rhoc(\pmb{s}) ( \ln \rhoc(\pmb{s}) -1) + 
			    \sum_\alpha  \rhopa(\pmb{s}) ( \ln \rhopa(\pmb{s}) -1) \right) \;.
	\eea
	The excess part is yet unknown. Its first functional derivative with respect to polymer densities is denoted by
	\bea
	   c^{(1)}_{\text{p},\alpha}(\pmb{s}) = - \beta \frac{\delta \FAO^\text{ex}[\rhoc, \{\rhopa\}]}{\delta \rhopa(\pmb{s})}\;,
	\eea 
	it is the first--order polymer direct correlation function which
	is needed to integrate out the polymers and to arrive at an effective AO functional for colloids only.
	This proceeds via the introduction of a semi--grand functional $\Upsilon$:
	\bea
	  \varUpsilon[\rho_\text{c}, \{ \rhopa \} ] &=& \FAO[\rhoc, \{\rhopa\}]  - \mup \sums \sum_\alpha \rhopa(\pmb{s})\;,  
	  \label{eq:semigrand}
	\eea
	whose minimization with respect to $\{\rhopa\}$ defines the effective AO functional:
	\bea
		\FAO^\text{eff}[\rhoc(\pmb{s}); \rhopr] &=& \text{min}_{\rhopa(\pmb{s})}  \varUpsilon[\rho_\text{c}, \{ \rhopa \} ]
        \eea
	The minimizing polymer density profile is given by
	\bea
	   \rhopa(\pmb{s}) = \rhopr \exp( c^{(1)}_{\text{p},\alpha}(\pmb{s}) ) \;. 
	\eea  
	The effective AO functional above is not yet the lattice gas functional $\Flg$ since it contains an additional 
	constant and one--body term in $\rhoc$ which must be subtracted:
	\bea
	 \label{eq:flg_sub}
	{
	\Flg[\rhoc(\pmb{s}); \epsilon] = \FAO^\text{eff}[\rhoc(\pmb{s}); \rhopr] + \rhopr \sums \sum_\alpha 
					   \left( 1- \sum_{\pmb{s}' \in \{\pmb{s},\pmb{s}+\hat{\pmb{e}}_\alpha\}} \rhoc(\pmb{s}') \right) \;.}
	\eea
	Here, the sum over $\pmb{s}'$ extends over lattice site $\pmb{s}$ and its nearest neighbor in $\alpha$--direction.
	The problem of finding the lattice gas functional $\Flg$ is thus shifted to finding the excess functional for
	the colloid--polymer mixture $\FAO^\text{ex}[\rhoc, \{\rhopa\}]$. For that, one might be tempted to apply
	the so--called ``linearization trick'', known from continuum models
	\cite{Brader2003} and applicable also to lattice models \cite{Mortazavifar2017}: One starts out with an excess functional for a hard mixture of
	colloids and rods of length 2 (derived from FMT) and linearizes with respect to the rod densities. (The rod--rod
	second--order direct correlation function is zero in the linearized functional as one would assume for polymeric,
	ideal--gas rods.) However, in that way one only recovers the mean--field functional, Eq.~(\ref{eq:fmf}), for the lattice gas
	\cite{Mortazavifar2017}. 

	Therefore we adopt a different approach, introduced under the name ``Highlander'' functional in Ref.~\cite{Cuesta2005}. 
	The polymers are ideal gas particles and thus each lattice site might be occupied by more than one polymer.
	For each of the polymer species, we define clusters of $n$ polymers occupying the same lattice site as being a separate cluster species.
	Particles of these new cluster species mutually interact with hard--core interactions: putting a cluster of $n$ polymers on top
	of a cluster of $m$ polymers results in a cluster of $n+m$ polymers, being a different species. In that way, we transform the
	colloid--polymer mixture into a mixture of colloids and hard polymer clusters which we will treat further with lattice FMT methods.  

	\subsubsection{Ideal gas polymer free energy in terms of polymer clusters} 

	Let $\Npa$ be the number of polymers of species $\alpha$ and $\Npcan$ the number of polymer clusters of species $\alpha$, containing $n$ polymers.
	The canonical partition function of a homogeneous $\alpha$--polymer system is given by
	\bea
	   Z^{(\text{p},\alpha)} = \frac{1}{\Npa !} |\Lambda|^\Npa \;,
	\eea
	where $|\Lambda|$ denotes the number of lattice sites of a particular lattice $\Lambda$. The associated grand partition function is defined  by
	\bea
	  \Xi^{(\text{p},\alpha)} = \sum_{\Npa=0}^\infty e^{\beta \mupa {\Npa}} Z^{(\text{p},\alpha)} \;. 
          \label{eq:gpf_pol}
	\eea
	As shown in App.~\ref{app:A}, this grand partition function can be written in terms of a polymer cluster partition function as follows:
	\begin{equation}
	\Xi^{(\text{p},\alpha)}= \sum_{ \{N_{\text{pc}^k,\alpha }\}_k }\;
    \prod_{m=1}^{\infty} \left( e^{\beta \mu_{\text{pc}^m,\alpha} N_{\text{pc}^m,\alpha}} \right)  \tilde{Z}^{(\text{pc}_\alpha^n)_n }(\{N_{\text{pc}_\alpha^n}\}_n, \Lambda)  \frac{1}{ \prod_{n}n!^{^{\Npcan}} }\;.
    \label{eq:xi_polymer}
    \end{equation}
	Here, $\{ N_{\text{pc}^k,\alpha }\}_k$ is the set of particle numbers of $\alpha$--polymer clusters with size $k=\{1,...,\infty\}$, and the sum over it means a sum from zero to infinity for each $N_{\text{pc}^k,\alpha}$ in this set. The polymer cluster chemical potential is given by
	\bea
		\mupcan = n \mupa \;.
		\label{eq:mupc}
	\eea
	The canonical polymer cluster partition function is given by
	\bea
		\tilde Z^{(\text{pc}^n,\alpha)}( \{\Npcan \} ) = \frac{1}{\left(\prod_n \Npcan !\right)} \frac{ |\Lambda|!}{\left(  |\Lambda| - \sum_n \Npcan  \right)!}
		\label{eq:z_pc}
	\eea 
	and is seen to correspond to a partition function of a multicomponent 
	lattice gas (composed of these $n$--polymer clusters) where the particles interact via site exclusion. Using Eq.~\eqref{eq:fexhc}, we can thus express the ideal gas free energy functional of the
	$\alpha$--polymers as
	\bea
	  \beta \Fid [ \rhopa] &=& \beta \Fid [ \{ \rhopcan \}  ] + {\sum_{\pmb{s}}}\pzd\left( \sum_n \rhopcan(\pmb{s}) \right) \\
	     \beta \Fid [ \{ \rhopcan \}  ] &=& \sums \left( \sum_n \rhopcan(\pmb{s}) 
		 ( \ln \rhopcan(\pmb{s}) -1)  + \sum_n \ln n! \, \rhopcan(\pmb{s})     \right) \;. \label{eq:fid_pc}
	\eea
	Here, the density $\rhopcan(\pmb{s})$ is the local density of $n$--polymer clusters of species $\alpha$.  
	According to this, the ideal gas polymer free energy is a sum of an ideal gas free energy of polymer clusters and 
	an excess free energy for a multicomponent lattice gas {with no attractions} (see Eq.~(\ref{eq:fexhc})). The
	ideal gas free energy of polymer clusters contains the standard multicomponent ideal gas free energy
	and a term which takes care of the last combinatorial factor appearing in Eq.~(\ref{eq:xi_polymer}).

	The free energy for the total polymer system is then the sum of this free energy over all polymer species:
	\bea
	  \beta \Fid [ \{ \rhopa \} ] &=& \sum_\alpha \beta \Fid [ \rhopa] \;.
	\eea  

	\subsubsection{Excess free energy functional of polymer clusters and colloids}

	\begin{figure}
	  \centerline{\includegraphics[width=8cm]{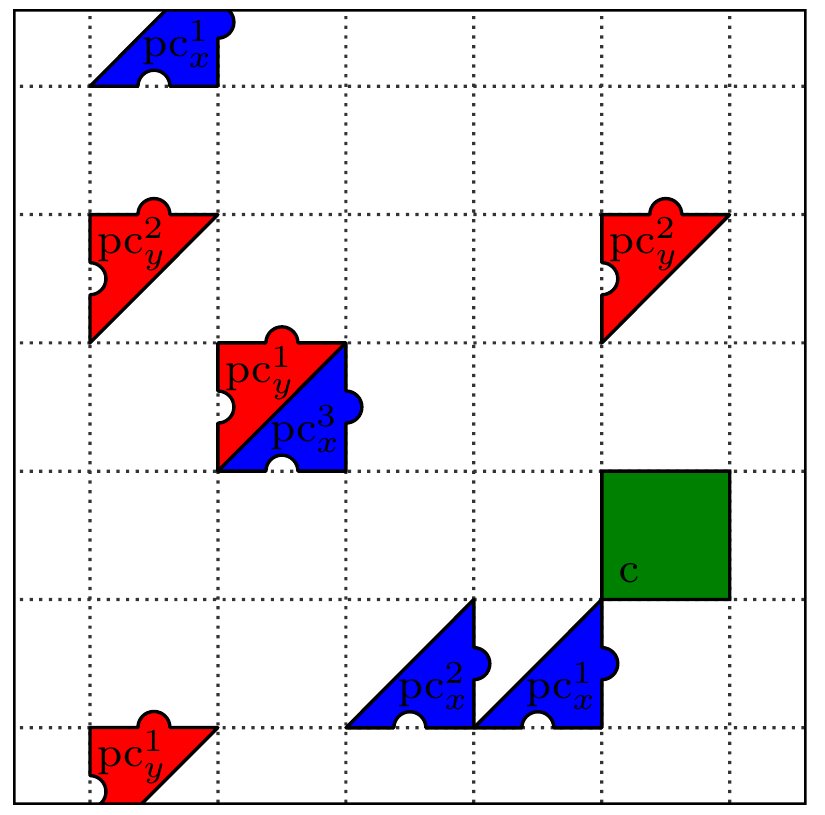}}
	  \caption{Visualization of the mixture of colloids and polymer clusters in 2D. Any mutual overlap of (parts of) the different
	shapes are forbidden.}
	  \label{fig:pcdef}
	\end{figure}

	For a visualization of the colloid--polymer cluster mixture with their hard interactions in 2D, see Fig.~\ref{fig:pcdef}.
	
	The free energy functional of the AO mixture of colloids and polymer clusters is split according to
	\bea
	  \FAO[\rhoc, \{\rhopa\}] \to \tFAO[\rhoc, \{\rhopcan\}] &=& \Fid[\rhoc] + \Fid [ \{ \rhopcan \}  ] + \tFAO^\text{ex}[\rhoc, \{\rhopcan\}] \label{eq:fAO tilde}
	\eea
	where the first term is the ideal gas free energy of colloids and the second term is the
	free energy of all polymer clusters (see Eq.~(\ref{eq:fid_pc})). Therefore $\tFAO^\text{ex}$ is different
	from $\FAO^\text{ex}$ in Eq.~(\ref{eq:fao}). 
	This new excess part $\tFAO^\text{ex}$ is constructed using the lattice FMT recipe of Lafuente and Cuesta \cite{Lafuente2002,Lafuente2004}. It can be briefly described as follows:
	Find the set of maximal 0D cavities and iteratively construct a free energy density which makes exact the free energy of those cavities.
	This results in the lattice FMT functional density.
	Here, a 0D cavity consists of a set of lattice points for each species with the following property: If one particle of a certain species occupies
	one of the points in the set, no other particle will fit in the cavity. The 0D cavity is maximal if no further points can be added to the set.   

	The whole construction is explained in App.~\ref{app:B} and here we just state the results. For the 2D lattice gas we find
	\bea
	  \beta  \tFAO^\text{ex}[\rhoc, \{\rhopcan\}] &=& \sums \Phi_\text{AO,2D}(\pmb{s}) \\
	  \Phi_\text{AO,2D}(\pmb{s}) &=& \sum_{i=1}^4 \pzd(n_i(\pmb{s})) - \sum_{i=5}^6 \pzd(n_i(\pmb{s})) - 3 \pzd(n_7(\pmb{s})) \nonumber \;.
	  \label{eq:Phi2D}
	\eea 
	The weighted densities $n_1 ...n_7$ are defined by
	\bea
	   n_1(\pmb{s})= &\rho_{\text{pc},x}(\pmb{s})+\rho_\text{c}(\pmb{s}), & n_5(\pmb{s})=\rho_{\text{pc},x}(\pmb{s}), \nonumber \\
	   n_2(\pmb{s})= &\rho_{\text{pc},x}(\pmb{s})+\rho_\text{c}(\pmb{s}+\hat{\pmb{e}}_x),\qquad & n_6(\pmb{s})=\rho_{\text{pc},y}(\pmb{s}), \nonumber \\
	   n_3(\pmb{s})= &\rho_{\text{pc},y}(\pmb{s})+\rho_\text{c}(\pmb{s}), &                 n_7(\pmb{s})=\rho_{\text{c}}(\pmb{s}), \nonumber \\
	   n_4(\pmb{s})= &\rho_{\text{pc},y}(\pmb{s})+\rho_\text{c}(\pmb{s}+\hat{\pmb{e}}_y) \qquad & \label{eq:ndef}
	\eea 
	Here,
	\bea
	    \rhopca(\pmb{s}) = \sum_n \rhopcan(\pmb{s})
	\eea
	is the total density of polymer clusters of species $\alpha$ at a lattice point and $\hat{\pmb{e}}_\alpha$ are translations of one lattice site in 
	$\alpha$--direction.

	For the 3D lattice gas we find
	\bea
	  \beta  \tFAO^\text{ex}[\rhoc, \{\rhopcan\}] &=& \sums \Phi_\text{AO,3D}(\pmb{s}) \\
	  \Phi_\text{AO,3D}(\pmb{s}) &=& \sum_{i=1}^6 \pzd(m_i(\pmb{s})) - \sum_{i=7}^9 \pzd(m_i(\pmb{s})) - 5 \pzd(m_{10}(\pmb{s})) \nonumber \;.
	  \label{eq:Phi3D}
	\eea 
	The weighted densities $m_1 ...m_{10}$ are defined by
	\bea
	   m_1(\pmb{s})= &\rho_{\text{pc},x}(\pmb{s})+\rho_\text{c}(\pmb{s}), &                 m_7(\pmb{s})=\rho_{\text{pc},x}(\pmb{s}), \nonumber \\
	   m_2(\pmb{s})= &\rho_{\text{pc},x}(\pmb{s})+\rho_\text{c}(\pmb{s}+\hat{\pmb{e}}_x),\qquad & m_8(\pmb{s})=\rho_{\text{pc},y}(\pmb{s}), \nonumber \\
	   m_3(\pmb{s})= &\rho_{\text{pc},y}(\pmb{s})+\rho_\text{c}(\pmb{s}), &                 m_9(\pmb{s})= \rho_{\text{pc},z}(\pmb{s}), \nonumber \\
	   m_4(\pmb{s})= &\rho_{\text{pc},y}(\pmb{s})+\rho_\text{c}(\pmb{s}+\hat{\pmb{e}}_y), \qquad&  m_{10}(\pmb{s})=\rho_{\text{c}}(\pmb{s}), \nonumber \\
	   m_5(\pmb{s})= &\rho_{\text{pc},z}(\pmb{s})+\rho_\text{c}(\pmb{s}), & \label{eq:mdef} \nonumber \\
	   m_6(\pmb{s})= &\rho_{\text{pc},z}(\pmb{s})+\rho_\text{c}(\pmb{s}+\hat{\pmb{e}}_z) \qquad &
	\eea 

	It is important to note that the excess free energy densities $\Phi_\text{AO,2D}(\pmb{s})$ and $\Phi_\text{AO,3D}(\pmb{s})$ depend 
	only on the total polymer cluster densities $\rhopca$. Furthermore, they
	depend \textit{locally} on $\rhopca(\pmb{s})$ and \textit{non--locally} on the colloid density $\rhoc(\pmb{s})$ and $\rhoc(\pmb{s}+\hat{\pmb{e}}_\alpha)$.

	\subsubsection{The effective free energy functional}

	Equivalent to the treatment in Sec.~\ref{sec:effAO}, we obtain the effective functional (depending on the colloid density profile only)
	by forming a semi--grand functional (coupling the polymer clusters grand--canonically) and minimize with respect to the 
	polymer cluster densities. The semi--grand functional is given by
	\bea
	  \varUpsilon[\rho_\text{c}; \{ \rhopcan \} ] &=& \tFAO[\rhoc, \{\rhopcan\}]  - \mup \sums \sum_\alpha \rhopa(\pmb{s})  \nonumber\\
					 &=& \tFAO[\rhoc, \{\rhopcan\}]  - \sum_n (n \mup) \sums \sum_\alpha \rhopcan(\pmb{s}) \;. 
        \eea
	The minimization is with respect to the polymer cluster densities $\rhopcan$ and results in 
	\bea
		    \rhopcan(\pmb{s}) = \frac{(\rhopr)^n}{n!} \exp( c^{(1)}_{\text{pc},\alpha}(\pmb{s}) ) 
		    \label{eq:rhopcan}
	\eea
	where
	\bea
  	c^{(1)}_{\text{pc},\alpha}(\pmb{s}) = - \beta \frac{\delta \tFAO^\text{ex}[\rhoc, \{\rhopcan\}]}{\delta \rhopca(\pmb{s})}\;,
	\eea
	Here, remember that $\tFAO^\text{ex}$ depends only on $\rhopcan(\pmb{s})$ through the total cluster density
	$\rhopca(\pmb{s})$. Eq.~(\ref{eq:rhopcan}) can be summed to give the total cluster density
	\bea
		\rhopca(\pmb{s}) &=& \zeta \, \exp( c^{(1)}_{\text{pc},\alpha}(\pmb{s}) ) \label{eq:rhopc_min} \\
		 \zeta &=& \exp( \rhopr) -1 \;.		
	\eea
	Here, $\zeta$ is a reservoir density for all $\alpha$--polymer clusters.
	The effective AO free energy functional is obtained by evaluating $\varUpsilon$ with $\rhopcan(\pmb{s})$ from
	Eq.~(\ref{eq:rhopcan}) and reads
	\bea
		{\beta}\tFAO^\text{eff}[\rhoc(\pmb{s});\rhopr] &=& \sums \rhoc(\pmb{s}) ( \ln \rhoc(\pmb{s}) -1) +
	    \left. {\beta}\tFAO^\text{ex}[\rhoc, \{\rhopcan\}] \right|_{\rhopcan (\text{Eq.}~(\ref{eq:rhopcan}))}  + \nonumber\\
	 & &    \left. \sums \sum_\alpha \rhopca(\pmb{s}) \left( -1 + c^{(1)}_{\text{pc},\alpha}(\pmb{s})   \right) \right|_{\rhopcan (\text{Eq.}~(\ref{eq:rhopcan}))} \;.
	\eea
Note that the same effective AO free energy functional is obtained
	by minimizing the following semi--grand functional (where the sum of polymer clusters is coupled grand--canonically)
	\bea
	  \tFAO^\text{eff}[\rhoc(\pmb{s});\rhopr] &=& \text{min}_{\rhopca(\pmb{s})} \tilde \varUpsilon[\rho_\text{c}; \{ \rhopca \} ]\\
	  \beta \tilde \varUpsilon[\rho_\text{c}; \{ \zeta_\alpha \} ] &=& 
                      \beta \Fid[\rhoc] + \sum_\alpha \beta \Fid[\rhopca]
		      + \beta\tFAO^\text{ex}[\rhoc, \{\rhopca\}] \nonumber \\
		       & & - \ln \zeta   \sums \sum_\alpha \rhopca(\pmb{s}) \;.  \label{eq:Upst}
        \eea
%
Here, $\beta \Fid[\rhopca] =\sums \rhopca(\pmb{s}) ( \ln \rhopca(\pmb{s}) -1)$ is the standard ideal gas free energy, i.e. without the combinatorial term $\propto \ln n!$ which appeared in the ideal gas
free energy \eqref{eq:fid_pc} of the individual polymer clusters.
	Thus one sees that the infinite number of polymer cluster densities $\rhopcan$ is not needed (we need only
	their sum); we have reformulated the AO model in terms of as many polymer cluster species as there were
	polymer species in the original AO model. (Here we have notationally equated
	$\tFAO^\text{ex}[\rhoc, \{\rhopcan\}]\equiv \tFAO^\text{ex}[\rhoc, \{\rhopca\}]$ since the excess functional only
	depends on $\rhopca$.)

	Finally, the lattice gas functional is obtained by performing the same subtraction of a constant and one--body term
	as in Eq.~(\ref{eq:flg_sub}):
	\bea
         \Flg[\rhoc(\pmb{s}); \epsilon] = \tFAO^\text{eff}[\rhoc(\pmb{s}); \rhopr] + \rhopr \sums \sum_\alpha
                                           \left( 1 - \sum_{\pmb{s}' \in \{\pmb{s},\pmb{s}+\hat{\pmb{e}}_\alpha\}} \rhoc(\pmb{s}') \right) \;. 
     \label{eq:flg_final}
    \eea

    We keep the name ``Highlander functional'' for this functional, as introduced in Ref.~\cite{Cuesta2005} for 
    a lattice AO functional which makes use of polymer clusters.

	This Highlander functional in 2D and 3D is only an approximation and not exact, and it is instructive to 
	look at the reasons. In 1D, the construction for the mixture of lattice colloids and one polymer species
	had already been performed in Ref.~\cite{Cuesta2005} (see also App.~\ref{app:B}) and shown to be exact, and since this mixture is 
	equivalent to the lattice gas/Ising model, the functional (\ref{eq:flg_final}) is exact in 1D. 
	This property is based on the exactness of the lattice FMT construction in 1D for nonadditive mixtures
	of hard rods where the nonadditivity extends to one lattice site \cite{Lafuente2002}. However, lattice FMT is not exact anymore in 2D and 3D (even though for small rod lengths or particle extensions the results are very close to simulation results \cite{Gschwind2017}). As a consequence, further efforts in improving lattice FMT for hard particles would also help improving density functionals for the lattice gas/Ising model.

%% file: sec_3_results.tex
\section{Results}
\label{sec:results}

\subsection{Bulk properties and phase diagram}

A bulk state in the lattice gas is characterized by a constant colloid density $\rhoc$ at a temperature $T$, the latter fixes
the polymer reservoir density $\rhopr=\beta\epsilon$ and the reservoir density of polymer clusters $\zeta = \exp(\beta\epsilon)-1$.
The individual polymer cluster densities are the same, $\rhopca = \rhopc$, and constant in space.
The evaluation of Eq.~(\ref{eq:rhopc_min}) gives for all dimensions
\begin{equation}
                \rho_{\text{pc}}=\zeta \frac{(1-\rho_\text{c}-\rho_\text{pc})^2}{1-\rho_\text{pc}}\;, 
\end{equation}
and solving this for $\rho_{\text{pc}}$ one obtains
\begin{equation}
        \rho_{\text{pc}}(\rho_\text{c}, \, \zeta)=\frac{1}{2(\zeta+1)}\left\lbrace 2\zeta (1-\rho_\text{c})+1
        - \sqrt{4\zeta (1-\rho_\text{c})\rho_{\text{c}}+1}\right\rbrace \qquad.
	 \label{eq:rhopc_bulk}
\end{equation}
This solution with the minus sign in front of the square root is consistent with the physical condition
that if $\zeta \to 0$ (no attraction) then $\rhopc \to 0$ (no polymers).
The free energy density $f$ follows from Eq.~(\ref{eq:flg_final}) (using Eqs.~(\ref{eq:Upst}) 
and (\ref{eq:Phi1D},\ref{eq:Phi2D},\ref{eq:Phi3D}) for the different dimensions) and is given by 
\begin{eqnarray*}
	\text{(1D)} \qquad	\beta f(\rho_\text{c}) &=&
        \rho_{\text{c}}\left( \ln\rho_{\text{c}}-1\right)
        +\rho_{\text{pc}}\left(\ln\rho_{\text{pc}}-1 \right) \notag\\
	& &+2\pzd(\rho_{\text{c}}+\rho_{\text{pc}})
        -\pzd(\rho_{\text{pc}})
        -\pzd(\rho_{\text{c}})  \\
	& &+\ln(1+\zeta)(1-2\rho_{\text{c}}) -  \rhopc\ln\zeta  \notag\\
	\text{(2D)} \qquad \beta f(\rho_\text{c})  &=&
        \rho_{\text{c}}\left( \ln\rho_{\text{c}}-1\right)
        +2\rho_{\text{pc}}\left(\ln\rho_{\text{pc}}-1 \right) \notag\\
	& &+4\pzd(\rho_{\text{c}}+\rho_{\text{pc}})
        -2\pzd(\rho_{\text{pc}})
        -3\pzd(\rho_{\text{c}}) \notag \\
	& &+2\ln(1+\zeta)(1-2\rho_{\text{c}}) - 2\rhopc\ln\zeta   \\
	\text{(3D)} \qquad \beta f(\rho_\text{c}) &=&
        \rho_{\text{c}}\left( \ln\rho_{\text{c}}-1\right)
        +3\rho_{\text{pc}}\left(\ln\rho_{\text{pc}}-1 \right) \notag\\
	& &+6\pzd(\rho_{\text{c}}+\rho_{\text{pc}})
        -3\pzd(\rho_{\text{pc}})
        -5\pzd(\rho_{\text{c}}) \notag \\
	& &+3\ln(1+\zeta)(1-2\rho_{\text{c}}) -3 \rhopc\ln\zeta \qquad.
\end{eqnarray*}
where $\rhopc$ is given by Eq.~(\ref{eq:rhopc_bulk}).
 The chemical potential $\mu = d f/ d \rhoc$ reduces to the partial derivative
  $\mu = \partial f/ \partial \rhoc$ since the dependence on $\rhoc$ through 
  $\rhopc$ leads to $(\partial f / \partial \rhopc)\; (\partial \rhopc / \partial \rhoc) $ and
  $\partial f / \partial \rhopc=0$ (see Eq.~(\ref{eq:Upst})).
 Accordingly we find
\begin{eqnarray}
	\text{(1D)} \qquad
	\beta\mu(\rhoc)&=&\ln \rhoc + 2 (\pzd)'(\rhoc+\rho_{\text{pc}}) - (\pzd)'(\rhoc)-2\ln(1+\zeta) \\
	\text{(2D)} \qquad \label{eq:mu2d}
	\beta\mu(\rhoc)&=&\ln \rhoc + 4 (\pzd)'(\rhoc+\rho_{\text{pc}}) - 3(\pzd)'(\rhoc)-4\ln(1+\zeta)\\
\text{(3D)} \qquad \label{eq:mu3d}
	\beta\mu(\rhoc)&=&\ln \rhoc + 6 (\pzd)'(\rhoc+\rho_{\text{pc}}) - 5(\pzd)'(\rhoc)-6\ln(1+\zeta)
\end{eqnarray}
where $(\pzd)'(\eta)=\partial\pzd(\eta)/\partial \eta= -\ln(1-\eta)$. 
Note that for completeness, we have included the results of the 1D lattice gas which are exact, as discussed
before. 

	\begin{figure}
    \centering
    \includegraphics[width=.45\textwidth]{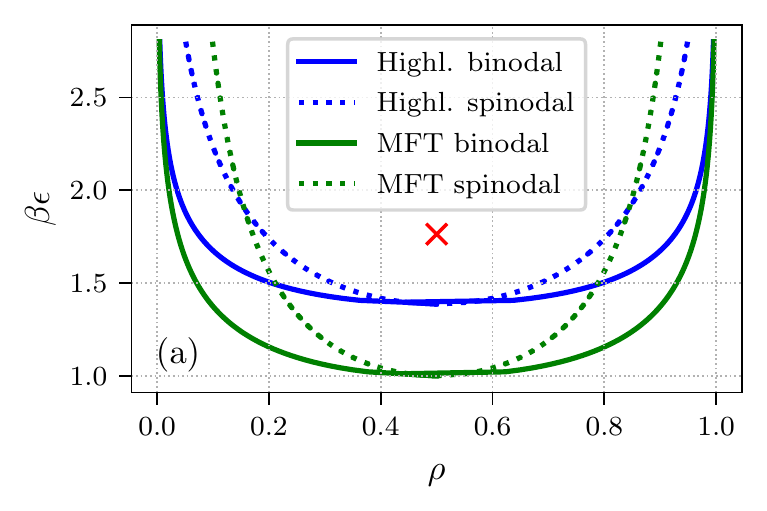}
    \includegraphics[width=.45\textwidth]{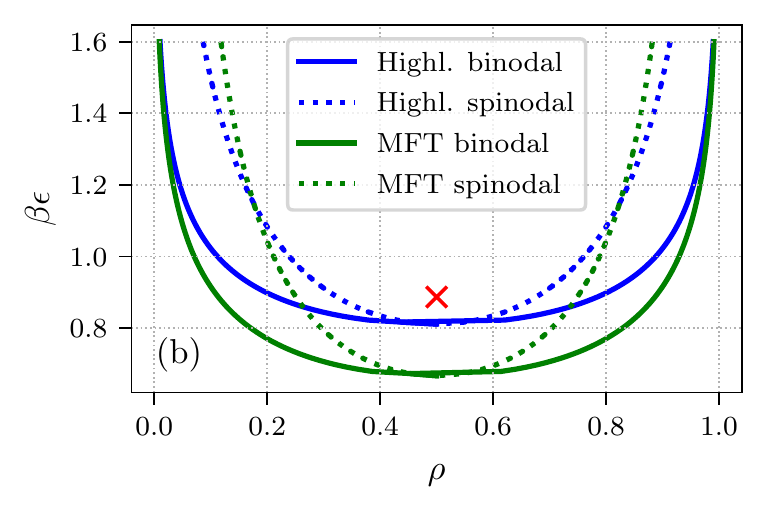}
    \caption{ 
    Binodals and spinodals of the lattice gas in (a) 2D and (b) 3D. ``MFT'' corresponds
		to the standard mean--field (Bragg--Williams) approximation and ``Highl.'' to the Highlander functional
		derived here. The red crosses mark (quasi--)exact critical points: in 2D from Onsager's solution \cite{Onsager1944} and in 3D from simulations \cite{Landau2018}.}
	  \label{fig:phasediagram}
\end{figure}

From the free energy density and the chemical potential, the pressure $p=\mu\rho-f$ can be calculated and the 
binodal can be evaluated using the Maxwell construction. The special symmetry of the lattice gas model entails that
the chemical potential at coexistence is given by:
\bea
  \mucoex = - \frac{z \epsilon}{2}
\eea
with $z$ being the number of nearest neighbours. Inserting this into Eqs.~(\ref{eq:mu2d},\ref{eq:mu3d}), one obtains an implicit equation for the coexisting densities
\begin{eqnarray}
  \text{(2D)} \qquad	0&\overset{!}{=}&\frac{\rhocoex(1-\rhocoex)^3}{(1-\rhocoex-\rho_\text{pc}(\rhocoex,\, \zeta))^4}-(\zeta+1)^2
        \label{binodal2d} \\
	\text{(3D)} \qquad     0&\overset{!}{=}&\frac{\rhocoex(1-\rhocoex)^5}{(1-\rhocoex-\rho_\text{pc}(\rhocoex,\, \zeta))^6}-(\zeta+1)^3
        \label{binodal3d}
\end{eqnarray}
The critical density is at $\rho_\text{crit}=0.5$ and
the critical temperature is at
\bea
	\text{(2D)} \qquad \ktc &=& \frac{1}{2 \ln 2}\, \epsilon \approx 0.72\, \epsilon  \\
	 \text{(3D)} \qquad \ktc &=& \frac{1}{2 \ln(3/2) }\, \epsilon \approx 1.23\, \epsilon
\eea
These are, incidentally, the critical temperatures of the Bethe--Peierls approximation \cite{Bethe1935}, as given in Ref.~\cite{Huang1987}. The full binodal and spinodal
(calculated by $\partial \mu/\partial\rhoc=0$) are shown in Fig.~\ref{fig:phasediagram}. Also the binodal is numerically equivalent to
the Bethe--Peierls binodal, although this is not obvious at all from the mathematical expressions leading to the
Bethe--Peierls approximation \cite{Huang1987}. It is a curious fact that we have an analytic form of the free energy density and
the equation of state equivalent leading to the Bethe--Peierls binodal. The equation of state, however,
does not agree with the equation of state obtained in Ref.~\cite{Pan1995} leading also to the Bethe--Peierls binodal.

\subsection{Planar interface}

\subsubsection{Numerical minimization}

The task is to perform the numerical minimization of $\Omega = \Flg - \sums \rhoc (\mu -\Ve)$ (with $\Flg$ given in
Eq.~(\ref{eq:flg_final})) with respect to $\rhoc(\pmb{s})$. However, the analytic derivative with respect to $\rhoc(\pmb{s})$
is quite involved, owing to the dependency of $\rhopc(\pmb{s})$ on $\rhoc(\pmb{s})$. It is therefore advisable
to minimize the  total grand functional
\bea
  \tilde \Omega [\rhoc, \{\rhopca\}] = \tilde \Upsilon - \sums \rhoc(\pmb{s}) (\mu -\Ve(\pmb{s})) {+ \rhopr \sums \sum_\alpha 
					   \left( 1- \sum_{\pmb{s}' \in \{\pmb{s},\pmb{s}+\hat{\pmb{e}}_\alpha\}} \rhoc(\pmb{s}') \right)}
\eea
(with $\tilde \Upsilon$ given in Eq.~(\ref{eq:Upst})) with respect to $\rhopc(\pmb{s})$ and $\rhoc(\pmb{s})$
simultaneously. The self--consistent equations for the colloid and polymer cluster density profiles
take a form suitable for Picard iteration. In 2D they read
\begin{eqnarray}
\rho_\text{c}(\pmb{s}) &=& z(\pmb{s})\, e^{4\rhopr}\, 
   \frac{\left( 1-n_1(\pmb{s})\right)\left(1-n_2(\pmb{s}-\hat{\pmb{e}}_x)\right)
   \left( 1-n_3(\pmb{s})\right)\left( 1-n_4(\pmb{s}-\hat{\pmb{e}}_y)\right) }{\left(1-n_7(\pmb{s})\right)^3} \notag\\
\rho_{\text{pc},x}(\pmb{s}) &=& \zeta\; \frac{\left( 1-n_1(\pmb{s})\right)\left(1-n_2(\pmb{s})\right) }{\left(1-n_5(\pmb{s})\right)} \notag\\
\rho_{\text{pc},y}(\pmb{s}) &=& \zeta\; \frac{\left( 1-n_3(\pmb{s})\right)\left(1-n_4(\pmb{s})\right) }{\left(1-n_6(\pmb{s})\right)} \qquad,\label{eq:min2d}
\end{eqnarray}
and in 3D
\begin{eqnarray}
\rho_\text{c}(s) &=& z(\pmb{s})\, e^{6\rhopr}\,
\frac{\splitfrac{\left(1-m_1(\pmb{s})\right)\left(1-m_2(\pmb{s}-\hat{\pmb{e}}_x)\right) \left( 1-m_3(\pmb{s})\right)\left( 1-m_4(\pmb{s}-\hat{\pmb{e}}_y)\right)}{\cdot \left(1-m_5(\pmb{s})\right)\left( 1-m_6(\pmb{s}-\hat{\pmb{e}}_z)\right)}}{\left(1-m_{10}(\pmb{s})\right)^{5}} \notag\\
\rho_{\text{pc},x}(s) &=& \zeta\;\frac{\left( 1-m_1(\pmb{s})\right)\left(1-m_2(\pmb{s})\right) }{\left(1-m_7(\pmb{s})\right)} \notag\\
\rho_{\text{pc},y}(s) &=& \zeta\;\frac{\left( 1-m_3(\pmb{s})\right)\left(1-m_4(\pmb{s})\right) }{\left(1-m_8(\pmb{s})\right)} \notag\\
\rho_{\text{pc},z}(s) &=& \zeta\;\frac{\left( 1-m_5(\pmb{s})\right)\left(1-m_6(\pmb{s})\right) }{\left(1-m_9(\pmb{s})\right)}\qquad.
\label{eq:min3d}
\end{eqnarray}
For both sets of equations, $z(\pmb{s})=\exp(\beta[\mu -\Ve(\pmb{s})])$. The definitions of the weighted densities $n_i(\pmb{s})$
are given in Eq.~(\ref{eq:ndef}) and the definitions of the $m_i(\pmb{s})$ is given in Eq.~(\ref{eq:mdef}). The Picard iterations are done in a standard manner with suitable mixing of old and new density profiles.

\begin{figure}
    \centering
    \includegraphics[width=.45\textwidth]{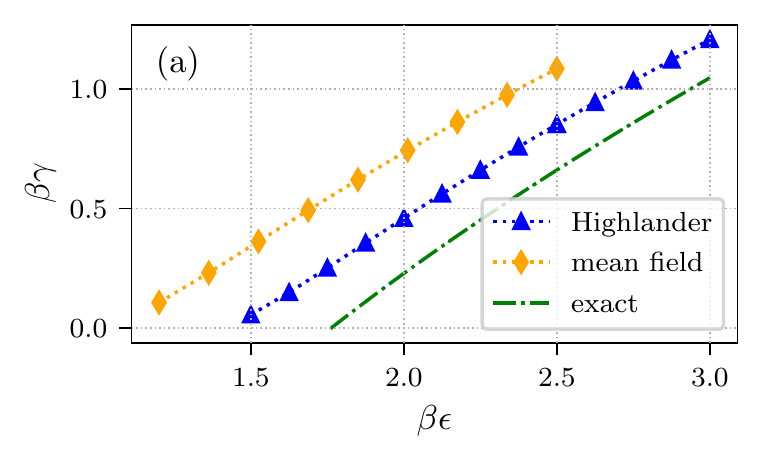}
    \includegraphics[width=.45\textwidth]{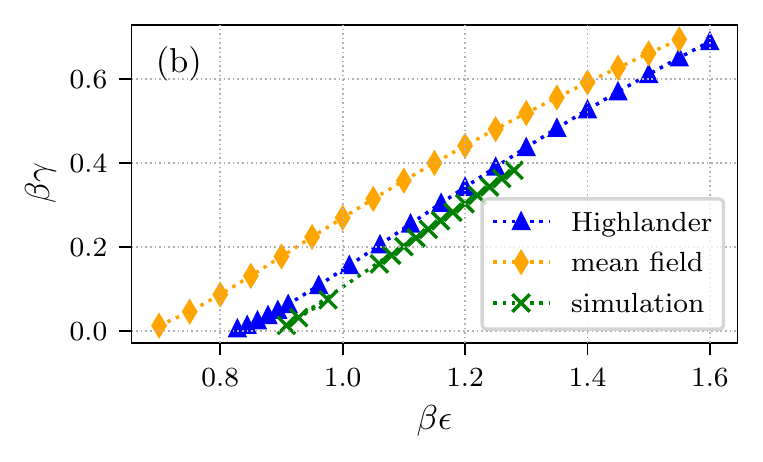}
    \caption{
    Surface tension $\beta\gamma$ of the free liquid-vapor interface as a function of inverse temperature, obtained
    with a box length $M=30$: (a) two dimensions and (b) three dimensions. Diamond symbols are results from the mean--field functional
    and triangles from the Highlander functional. The green dot--dashed line in (a) shows Onsager's exact results \cite{Onsager1944} and
    the cross symbols in (b) are simulation results from Refs.~\cite{Berg1993} and \cite{Bittner2009}.}
    \label{fig:gamma-epsi@mu_coex}
\end{figure}

\subsubsection{Free liquid-vapor interface}
\label{sec:freeinterface}
We consider interfaces in 2D with orientation [10] and interfaces in 3D with orientation [100]. The density profile varies in $x$--direction and is constant in the other direction(s).
We choose $\mu=\mucoex$, the chemical potential at coexistence.
We connect the system with a reservoir of bulk liquid and vapor at coexistence densities at the two $x$-boundaries of the computational box with length $M$ and upon iteration, a free liquid-vapor interface must evolve at some place within the system.\par
 At coexistence, the bulk liquid pressure $p_\text{l}$ and vapor pressure $p_\text{v}$ are the same and the surface tension
 is just the excess grand free energy density in the system:
\begin{equation}
    \gamma = \frac{\Omega}{A} + p_\text{v} M = \frac{\Omega}{A} + p_\text{l} M  \;,
\end{equation}
Here $\Omega/A$ is a line density in 2D and an aerial density in 3D. In computing $\gamma$ for the Highlander functionals (which employ nonsymmetric weighted densities), some subtleties arise which
are discussed in App.~\ref{app:C}.

Figure~\ref{fig:gamma-epsi@mu_coex} shows the surface tension of the free liquid-vapor interface at coexistence 
as a function of inverse temperature, here the results from the mean--field functional (Eq.~\eqref{eq:fmf}) and the Highlander functional are compared
with exact values (in 2D) and simulation values (in 3D).
In two dimensions the exact surface tension is given by \cite{Onsager1944}
\begin{equation}
    \beta\gamma = \frac{\beta\epsilon}{2}-\log\left(\coth\left(\frac{\beta\epsilon}{4}\right)\right)\;.
\end{equation}
The Highlander functional systematically improves the mean--field values,  and in 3D the Highlander results are close to the simulation
results. However, in 3D near the critical point the Highlander results show typical mean-field behavior ($\gamma \propto (T_c-T)^{1.5}$) which
is different from the exact Ising behavior ($\gamma \propto (T_c-T)^{1.26...}$).

\subsubsection{Constrained liquid--vapor interface}
\label{sec:constrainedinterface}

As before, we connect the system with a reservoir of bulk liquid and vapor 
but we look for solutions with a fixed average density (fixed number of particles) in the system.
This necessitates the minimization of the free energy functional $\Flg$ with this constraint which is
implemented by treating the chemical potential $\mu$ as a Lagrange multiplier.
In continuum models one would find $\mu=\mucoex$, and (almost) any value for a fixed average density between $\rho_\text{v}$
and  $\rho_\text{l}$ can be realized through a shift of the interface position in the box. 
(This is a consequence of the translational symmetry in continuum models.)
The interface position $\xem$ can be defined with the
equimolar condition which for a continuum model reads
\bea
  \int_0^{\xem} (\rho_\text{l} - \rho(x)) dx &= & \int_{\xem}^M (\rho(x) - \rho_\text{v})   dx \quad \rightarrow \\
  \xem &=& \frac{1}{\rho_\text{l} - \rho_\text{v} } \left(\int_0^M \rho(x) dx  - M \rho_\text{v}   \right)  
 \label{eq:xem}
\eea
\begin{figure}
    \centering
    \begin{subfigure}[b]{0.44\textwidth}
        \centering
        \includegraphics[width=\textwidth]{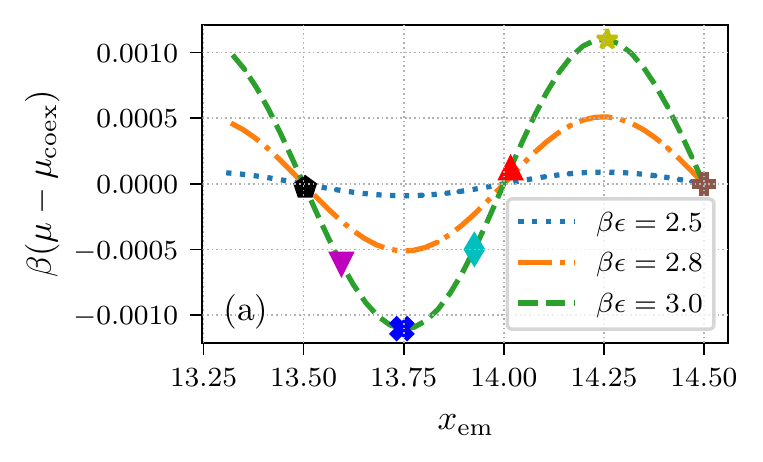}
    \end{subfigure}
    \qquad
    \begin{subfigure}[b]{0.44\textwidth}
        \centering
        \includegraphics[width=\textwidth]{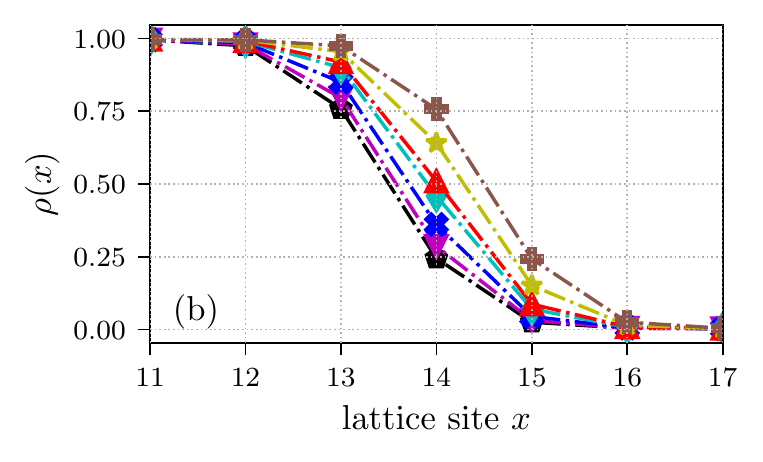}
    \end{subfigure}
    \qquad
    \begin{subfigure}[b]{0.44\textwidth}
        \centering
        \includegraphics[width=\textwidth]{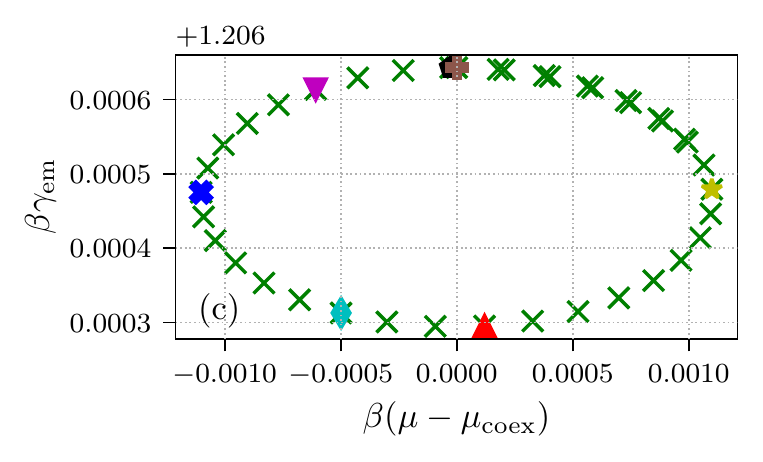}
    \end{subfigure}
    \caption{
    Properties of planar interfaces in 2D with constrained equimolar position $\xem$ from the Highlander functional. (a) Chemical potential as a function of $\xem$ for three inverse temperatures. For $\beta\epsilon=3.0$ and corresponding to the symbols, interface density profiles are shown in (b). 
    (c) Planar interface tension as function of chemical potential for $\beta\epsilon=3.0$, the respective
    symbols correspond to the systems marked with the same symbols in (a) and (b). System size is $M=30$.} 
    \label{fig:2d-highl@constrained surface}
\end{figure}
This condition can be used also for lattice models where we interpret the integral being evaluated with the trapezoidal rule on the 
discrete lattice. 
Here we use the convention that the numerical box extends from 0 to $M$ and the asymptotic state to the left is the 
liquid and the asymptotic state to the right is vapor. 
In the lattice gas model, the properties of the interface are a priori only invariant upon discrete shifts
of the interface, i.e. $\xem \to \xem^{(0)} + n$ where $\xem^{(0)}$ is the equimolar position of a free interface and $n$ is integer. Suppose that through the average density constraint one tries to put a few additional particles into
a system with a free interface. These additional particles can be accomodated by displacing the interface towards the vapor phase or they are accomodated in the bulk, i.e. they change the bulk densities. In general, we find that both mechanisms 
occur,  and we find solutions for arbitrary $[\xem] \neq [\xem^{(0)}]  $ (where we use the notation
$[x]= x- \text{int}(x)$ and $\text{int}(x)$ is the integer part of $x$) at 
chemical potentials $\mu(\xem) \neq \mucoex$, i.e., these are solutions for a planar interface off-coexistence.
We may define the interface tension associated with these off--coexistence solutions as
\begin{equation}
    \gem([\xem]) := \frac{\Omega}{A} + p_\text{v} M + \Delta p\; \xem \;. \label{eq:def_gamma}
\end{equation}
Here, we defined $\Delta p:=p_\text{l}(\mu)-p_\text{v}(\mu)$ which is in general not zero. 
The linear extension of the liquid phase is given by $\xem$ in our conventions. One sees that here the surface tension
depends on the arbitrary choice of the interface position, and the equimolar condition is only a convenient choice
for fixing it.

\begin{figure}
    \centering
    \includegraphics[width=0.45\textwidth]{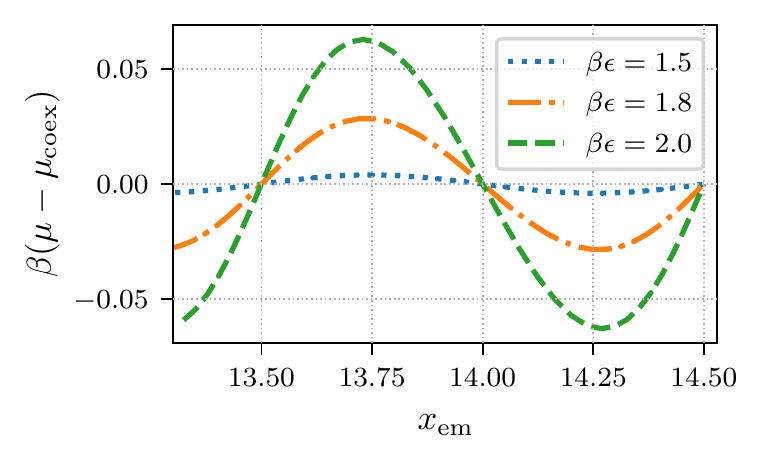}
    \caption{Chemical potential as function of the equimolar position in 2D for the mean--field functional, Eq.~\eqref{eq:fmf}. System size is $M=30$. }
    \label{fig:2d-mf@constrained surface}
\end{figure}

First we illustrate these off--coexistence solutions for the Highlander functional in 2D, see Fig.~\ref{fig:2d-highl@constrained surface}, for the three
inverse temperatures $\beta \epsilon=2.5,\;2.8$ and $3.0$ 
(we remind that the Highlander critical point is at $\beta_\text{c}\epsilon \approx 1.39$). As expected from the periodicity
of the lattice, $\mu(\xem)$ is a periodic function of $\xem$ with period 1 (see Fig.~\ref{fig:2d-highl@constrained surface}(a)). The amplitude of the oscillations in $(\mu-\mucoex)$ quickly decreases with raising the temperature, so
this is clearly an effect which is only relevant far away from the critical point. For the lowest temperature
$\beta\epsilon=3.0$ we show a few density profiles in Fig.~\ref{fig:2d-highl@constrained surface}(b)
which correspond to the states marked by symbols in  Fig.~\ref{fig:2d-highl@constrained surface}(a). 
The profiles with integer and half--integer values for $\xem$ are at coexistence and are symmetric upon interchanging
gas and liquid phase (an expected symmetry for the lattice gas model). Also one sees that by going from
$\xem=13.5$ (black pentagon) to $\xem=14.5$ (brown cross) the profile shifts by exactly one lattice site. All other profiles
are asymmetric. The associated surface tensions for these profiles are marked by the same symbols in a
plot of $\beta\gem$ as a function of $\beta(\mu-\mucoex)$ in Fig.~\ref{fig:2d-highl@constrained surface}(c). 
\begin{figure}
    \centering
    \begin{subfigure}[b]{0.44\textwidth}
        \centering
        \includegraphics[width=\textwidth]{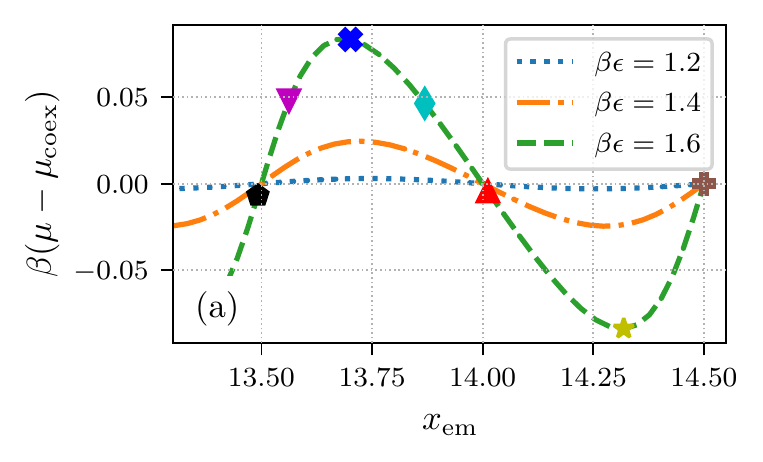}
    \end{subfigure}
    \qquad
    \begin{subfigure}[b]{0.44\textwidth}
        \centering
        \includegraphics[width=\textwidth]{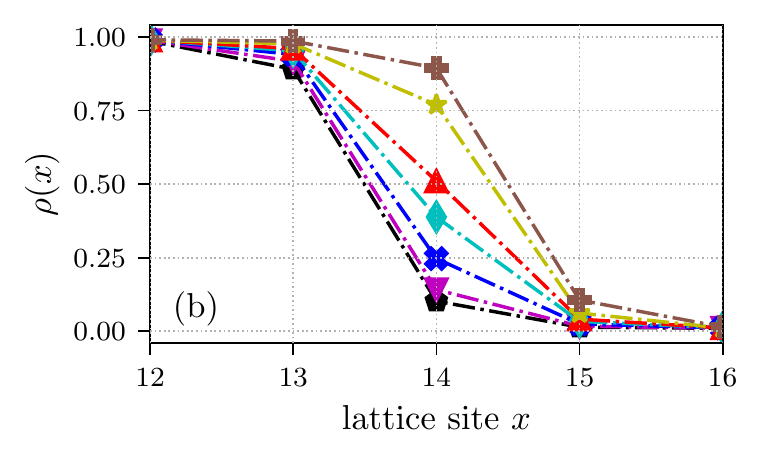}
    \end{subfigure}
    \qquad
    \begin{subfigure}[b]{0.44\textwidth}
        \centering
        \includegraphics[width=\textwidth]{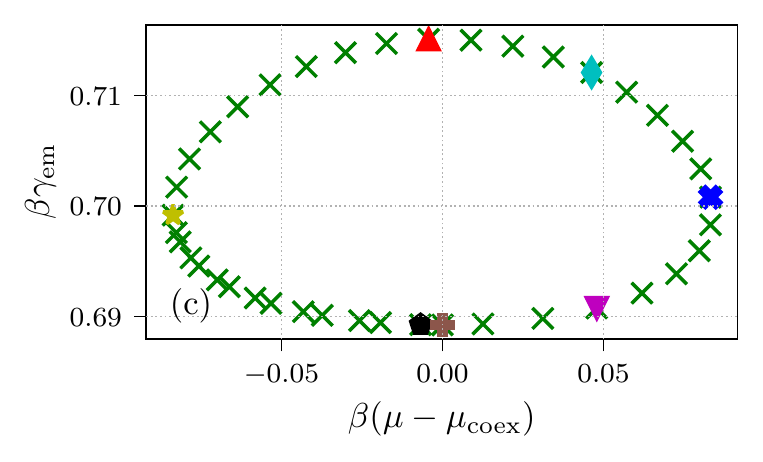}
    \end{subfigure}
    \caption{
    Properties of planar interfaces in 3D with constrained equimolar position $\xem$ from the Highlander functional. (a) Chemical potential as a function of $\xem$ for three inverse temperatures. For $\beta\epsilon=1.6$ and corresponding to the symbols, interface density profiles are shown in (b). 
    (c) Planar interface tension as function of chemical potential for $\beta\epsilon=1.6$, the respective
    symbols correspond to the systems marked with the same symbols in (a) and (b). System size is $M=30$.} 
    \label{fig:3d-highl@constrained surface}
\end{figure}

Here one sees that the profile with an integer value for $\xem$ (red triangle) gives a minimal surface tension
(in the actual numerics, it is difficult to fix $\xem$ precisely at an integer value, so the red triangle is
not exactly at the minimum). 
We checked that a free interface (as determined in Sec.~\ref{sec:freeinterface}) indeed gives $\xem$ at integer values.
The profiles with half--integer values for $\xem$ (black pentagon, brown cross) give a maximal surface tension. These results are actually somewhat counterintuitive. For $T \to 0$ ($\beta\epsilon \to \infty$) one expects a sharp interface (with a filled layer next to the empty gas phase) to be the state of
minimal free energy. Such an interface has an $\xem$ at half-integer value according to the definition
in Eq.~\eqref{eq:xem}. Raising the temperature implies broadening the interface by removing a few particles from the filled layer and building small terraces on it, but the associated $\xem$ would remain at a 
half--integer value. In contrast, an interface with $\xem$ at integer values requires
$\rho(\xem)=0.5$ and for $T\to 0$ that
would imply a half--filled interface layer next to the gas phase which should not be the equilibrium state.
Surprisingly, upon raising $\beta\epsilon$ the Highlander functional in 2D shows a transition of $\xem^{(0)}$ from 
half--integer to integer at around 2.19. So actually high temperature interfaces have an equimolar position as expected for zero temperature, and low temperature interfaces not.
We checked that until $\beta\epsilon=7.0$ the equimolar position stayed at integer value.
Presumably this is an artefact of the 2D Highlander functional.

As a comparison, in Fig.~\ref{fig:2d-mf@constrained surface} we show $\beta(\mu-\mucoex)$ as function of $\xem$ for the mean--field functional in 2D for the three inverse temperatures $\beta\epsilon=1.5,\;1.8$ and $2.0$
(here the critical point is at $\beta_\text{c}\epsilon = 1.0$). Here, the oscillations in the chemical
potential persist to {higher} temperatures. The corresponding free profiles at these temperatures have
half--integer values for $\xem^{(0)}$, so the appearance of off-coexistence interfaces does not seem to be tied with the peculiar occurrence of integer values for $\xem^{(0)}$.

\begin{figure}
    \centering
    \includegraphics[width=0.45\textwidth]{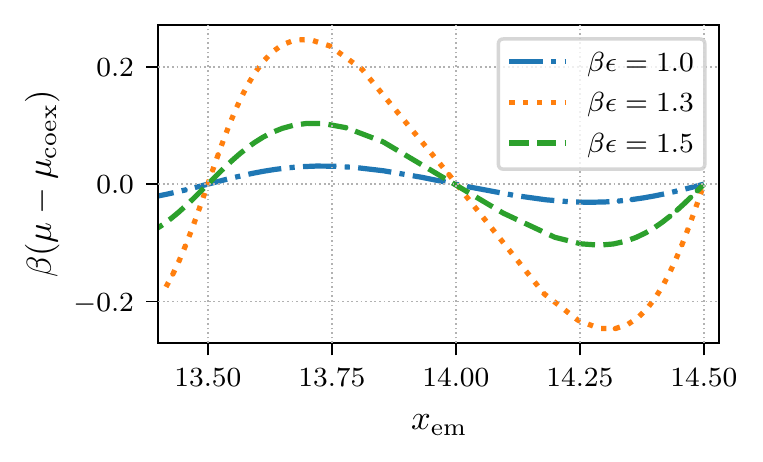}
    \caption{Chemical potential as function of the equimolar position in 3D for the mean--field functional, Eq.~\eqref{eq:fmf}. System size is $M=30$. }
    \label{fig:3d-mf@constrained surface}
\end{figure}

Turning to three dimensions, we find off--coexistence interfaces as well.
For the Highlander functional in 3D, see the results in Fig.~\ref{fig:3d-highl@constrained surface} for the three inverse temperatures $\beta \epsilon=1.2,\;1.4$ and $1.6$ (the Highlander critical point is at
$\beta_\text{c} \epsilon \approx 0.81$). The effect is actually bigger than in 2D (Highlander functional),
as can be seen from the amplitude of the oscillations in $\beta(\mu-\mucoex)$ (Fig.~\ref{fig:3d-highl@constrained surface}(a)) and the deviations of surface tensions from the 
equilibrium value of the free interface (Fig.~\ref{fig:3d-highl@constrained surface}(c)). For the
lowest temperature corresponding to $\beta\epsilon=1.6$ we show selected density profiles in
Fig.~\ref{fig:3d-highl@constrained surface}(b) in a similar way as in 2D. 
Again interfaces with integer and half--integer values for $\xem$ are at coexistence,
here the noticeable difference to the 2D case is that the interfaces with half--integer $\xem$ have minimal surface tension, i.e. are free interfaces.   

Similar to the 2D case, the mean--field functional in 3D produces stronger oscillations in chemical potential and surface tensions, and they persist for higher temperatures. This is illustrated in Fig.~\ref{fig:3d-mf@constrained surface} where we show $\beta(\mu-\mucoex)$ as function of $\xem$
for the three inverse temperatures $\beta\epsilon=1.0,\;1.3$ and $1.5$
(here the critical point is at $\beta_\text{c}\epsilon \approx 0.67$). One can conclude that the Highlander functional (although also of mean--field type) includes more correlations (or ``fluctuations'') which partly wash out the effect of the off--coexistence oscillations in interface properties.

To the best of our knowledge, the existence and properties of planar interfaces off--coexistence in the lattice gas or Ising model has
not been addressed by simulations before. A suitable method is the calculation of interface tensions via
the density of states $g(E,N)$ where $E$ is the total energy and $N$ the number of particles \cite{dePablo2003}. For a lattice--$\Phi^4$ model, Tr\"oster \textit{et al.} find oscillations
in the free energy per particle as function of continuous order parameter at low temperatures \cite{Troester2005a}. These
oscillations should correspond to the oscillating surface tension found here, and they also vanish for higher temperature, similar to our findings. In the end, the oscillations can be traced back to lattice peculiarities in the density of states which for the 2D Ising model is discussed more in detail in Ref.~\cite{Troester2005}.

%% file: sec_4_outlook.tex
\section{Conclusions and outlook}
\label{sec:outlook}

We have derived a density functional for the lattice gas using the idea of treating attractions as depletion interactions in a lattice colloid--polymer mixture and subsequently using methods from lattice fundamental measure theory. The resulting functional shows a significant improvement for the phase diagram when compared
to the standard mean field (Bragg--Williams) approximation, its binodal is equivalent to the Bethe--Peierls approximation. We have investigated planar interfaces in detail. Also surface tensions of free interfaces show a significant 
improvement over the standard mean--field functional, especially in 3D they are very close to simulation results. When the position of the interface is constrained to continuous values
(through a constraint on the average density in the system),
we have found planar interface solutions off--coexistence which implies that the asymptotic bulk states connected by the interface are at
different pressure. The effect becomes more pronounced at lower temperatures. Perhaps this effect is not only an artefact of a simplified lattice model but can also be seen in continuum models, e.g. when considering the planar interface between two solids at coexistence. 

The results appear to have some relevance for the droplet evaporation transition in the lattice gas model
and which is a suitable route to determine interface tensions of droplets \cite{Troester2017}.
This problem is currently under investigation. Also, from a systematic point of view it seems to be worthwhile to continue to work on the solution for the general inhomogeneous equilibrium problem in the lattice gas model which is unquestionably a simple yet very basic model for our understanding of statistical systems.
Building on our approach, further progress could be achieved by improving the lattice FMT functionals for hard particles
in two and three dimensions or by using novel machine learning techniques addressed to obtain analytic functionals \cite{Lin2020}. 

%% file: app_A_polymerclusters.tex
  \section{Mapping the partition function of polymers to a  partition function of polymer clusters}
\label{app:A}

Here we show that the grand partition function of a species of lattice polymers (Eq.~(\ref{eq:gpf_pol}))
is equivalent to the (slightly modified) grand partition function of a hard lattice gas of polymer clusters 
(Eqs.~(\ref{eq:xi_polymer},\ref{eq:mupc},\ref{eq:z_pc})).

We start by considering the partition function of $N_{\text{p},\alpha}$ ideal $\alpha$--polymers with no other components. 
The partition function is given by
\begin{equation}
	Z^{(\text{p},\alpha)}(N_{\text{p},\alpha}, \Lambda)
=\frac{1}{N_{\text{p},\alpha}!} \left| \Lambda\right| ^{N_{\text{p},\alpha}} \qquad, \label{partition function pi 1}
\end{equation}
where $\left| \Lambda\right| $ is the number of lattice sites of the lattice $\Lambda$. 
If a single lattice site is simultaneously occupied by $n$ polymers, 
we refer to it as an $n$--polymer cluster of species $\alpha$, short $\text{pc}^n,\alpha$. 
Every single configuration of the one--component polymer system can be considered as an arrangement of these clusters. 
Let $N_{\text{pc}^n,\alpha}$ be the number of stacks of size $n$, then there are
\begin{itemize}
	\item $\left| \Lambda\right| !/(\left| \Lambda\right| -\sum_{n}N_{\text{pc}^n,\alpha})!$ possibilities to distribute $\sum_{n}N_{\text{pc}^n,\alpha}$ distinguishable stacks on the lattice.
	\item $\prod_n N_{\text{pc}^n,\alpha}!$ multiple counts because stacks of the same size are indistinguishable.
	\item $N_{\text{p},\alpha}!/\prod_{n}n!^{N_{\text{pc}^n,\alpha}}$ possibilities to distribute the polymers on the stacks (exchanging two polymers in the same stack does not lead to a new configuration).
\end{itemize}
Using these factors, we thus may write the partition function of equation \eqref{partition function pi 1} by
\begin{align}
    Z^{(\text{p},\alpha)}(N_{\text{p},\alpha}, \Lambda)&=\frac{1}{N_{\text{p},\alpha}!} 
\left. \sum_{\{N_{\text{pc}^k,\alpha}\}_k} \right|_{\sum_{k}k N_{\text{pc}^k,\alpha}=N_{\text{p},\alpha}} \frac{\left| \Lambda\right| !}{\left(\left| \Lambda\right| -\sum_{n}N_{\text{pc}^n,\alpha}\right)!} \frac{1}{\left( \prod_n N_{\text{pc}^n,\alpha}!\right) } \frac{N_{\text{p},\alpha}!}{\left( \prod_{n}n!^{N_{\text{pc}^n,\alpha}}\right) } \notag\\
	&= 
\left. \sum_{\{N_{\text{pc}^k,\alpha}\}_k} \right|_{\sum_{k}k N_{\text{pc}^k,\alpha}=N_{\text{p},\alpha}}
\frac{1}{\left( \prod_n N_{\text{pc}^n,\alpha}!\right) } \frac{\left| \Lambda\right|! }{\left(\left| \Lambda\right| -\sum_{n}N_{\text{pc}^n,\alpha}\right)!}  \frac{1}{\left( \prod_{n}n!^{N_{\text{pc}^n,\alpha}}\right) } \;, & \label{partition function pi 2} 
\end{align}
where $
    \sum_{\{N_{\text{pc}^k,\alpha}\}_k} = \sum_{N_{\text{pc}^1,\alpha=1}}^\infty\dots\sum_{N_{\text{pc}^\infty,\alpha=1}}^\infty
$.
One should notice here that the factor
\begin{equation}
	\tilde{Z}^{(\text{pc}^n,\alpha)_n }(\{N_{\text{pc}^n,\alpha}\}, \Lambda):=\frac{1}{\left( \prod_n N_{\text{pc}^n,\alpha}!\right) } \frac{\left| \Lambda\right| !}{(\left| \Lambda\right| -\sum_{n}N_{\text{pc}^n,\alpha})!}
\end{equation}
is the partition function of the $n$--polymer cluster lattice gas where the particles interact via site exclusion (Eq.~(\ref{eq:z_pc})). 
Handling the constrained sum in Eq.~(\ref{partition function pi 2}) is done best in the grand ensemble 
where we can execute the summation over the number of particles from $0$ to $\infty$ for every species independently. 
With $\mu_{\text{p},\alpha}$ being the chemical potential of the polymers, the polymer fugacity can be split into the product
\begin{equation}
	z_{\text{p},\alpha} = e^{\beta \mu_{\text{p},\alpha} N_{\text{p},\alpha}}=\prod_{n=1}^{\infty}e^{\beta \mu_{\text{p},\alpha} n N_{\text{pc}^n,\alpha}}
\end{equation}
and the grand canonical partition function of the one component polymer system takes the form
\begin{equation}
	\Xi^{(\text{p},\alpha)}(T, \mu_{\text{p},\alpha}, \Lambda)=
	\sum_{\{N_{\text{pc}^k,\alpha}\}_k}\;
\prod_{m=1}^{\infty} \left( e^{\beta \mu_{\text{p},\alpha} m N_{\text{pc}^m,\alpha}} \right)  \tilde{Z}^{(\text{pc}^n,\alpha)_n }(\{N_{\text{pc}^n,\alpha}\}_n, \Lambda)  \frac{1}{ \prod_{n}n!^{N_{\text{pc}^n,\alpha}} }\;, \label{grand partition func 1 poly species}
\end{equation}
where the summation is over each $N_{\text{pc}^k,\alpha}$ in the set (for a given $\alpha$).
This is the grand partition function of a system composed of the infinite number of hard polymer clusters with the chemical potential 
for each cluster species given by $\mu_{\text{pc}^n,\alpha}:=n\mu_{\text{p},\alpha}$ but its partition function 
\begin{equation}
Z^{(\text{pc}^n,\alpha)_n }(\{N_{\text{pc}^n,\alpha}\}_n, \Lambda) := \tilde{Z}^{(\text{pc}^n,\alpha)_n }(\{N_{\text{pc}^n,\alpha}\}_n, \Lambda)  \frac{1}{\prod_{n}n!^{N_{\text{pc}^n,\alpha}}} \qquad. \label{partition function extended model}
\end{equation}
differs by the factor $1/\prod_{n}n!^{N_{\text{pc}^n,\alpha}}$ 
from $\tilde Z$.

It remains the question how the mapping to polymer clusters may be extended to a system of several (say, $d$) polymer components. 
Since all polymer are ideal particles, the partition function of a system composed of several polymer species is simply 
the product of every single species partition function \eqref{partition function pi 2}:
\begin{equation}
	Z^{\{(\text{p},\alpha)\}}(\{N_{\text{p},\alpha}\}_\alpha, \Lambda) = \prod_{\alpha=1}^{d} Z^{(\text{p},\alpha)}(N_{\text{p},\alpha}, \Lambda)
	\qquad.
\end{equation}
The grand partition function of the multi-polymer system then becomes
\begin{equation}
	\Xi^{\{(\text{p},\alpha)\}}(T, \{\mu_{\text{p},\alpha}\}_\alpha, \Lambda)= 
\sum_{\{N_{\text{pc}^k,\tau}\}_{k,\tau}}
	\prod_{\nu=1}^{d} \prod_{m=1}^{\infty} \left( e^{\beta \mu_{\text{p},\nu} m N_{\text{pc}^m,\nu}}\right)
	\prod_{\alpha=1}^{d} Z^{(\text{pc}^n,\alpha)_n }(\{N_{\text{pc}^n,\alpha}\}_n, \Lambda) \;.
\label{eq:Xi1}
\end{equation}
Note that here the sum over $\{N_{\text{pc}^k,\tau}\}_{k, \tau}$ runs from $0$ to $\infty$ for every $N_{\text{pc}^k,\tau}$ with $k=\{1,...,\infty\}$ and $\tau=\{1,...,d\}$.
Equation~(\ref{eq:Xi1}) has the form of a grand partition function of a system of an infinite number of polymer cluster species 
$\{\text{pc}^n,\alpha\}_{\alpha,n}$ determined by the canonical partition function
\begin{equation}
	Z^{(\text{pc}^n,\alpha)_{\alpha,n}} (\{N_{\text{pc}^n,\alpha}\}_{\alpha,n}, \Lambda) := 
\prod_{\alpha=1}^{d} Z^{(\text{pc}^n,\alpha)_{n}} (\{N_{\text{pc}^n,\alpha}\}_{n}, \Lambda) \qquad.
\end{equation}
From the exponential in Eq.~(\ref{eq:Xi1}) one reads off
the chemical potential of an $n$--polymer cluster of species $\alpha$ ($\text{pc}^n,\alpha$): 
$\mu_{\text{pc}^n,\alpha}=n\mu_{\text{p},\alpha}$. 
Therefore the multispecies cluster reformulation led to a model
where polymer clusters of the same species which only differ in the stack size interact via site exclusion, 
while polymer clusters from different polymer species do not influence each other, thus behave ideal. 
Thus it is intrinsically a non--additive model.

%% file: app_B_construction_excess.tex
  \section{Construction of the excess free energy functional $\tFAO^\text{ex}[\rhoc, \{\rhopcan\}]$}
\label{app:B}

The construction of lattice FMT functionals proceeds via the following iterative procedure \cite{Lafuente2004}. 
First, one finds a maximal set of 0D cavities.  A 0D cavity consists of a set of lattice points
for each species with the following property: If one particle of a certain species occupies one
of the points in the set, no other particle will fit in the cavity. The 0D cavity is maximal if
no further points can be added to the set. Pictorially, one can visualize these cavities with suitable
hard walls (as will be done below).
The requirement on the excess functional is that it gives 
the exact 0D free energy for a density distribution, compatible with any such maximal cavity at an arbitrary location.  
Second, the iterative procedure is started as follows: The excess free energy is a sum over
the 0D free energies of all such density distributions. However, when a specific cavity of the maximal set is evaluated with this trial 
functional, it will generate the correct 0D free energy plus some residual terms. All these residual terms are explicitly subtracted
in an updated excess free energy. Re--evaluation with a specific cavity may result in further residual terms which need to be subtracted again.
As shown in Ref.~\cite{Lafuente2005}, this procedure is guaranteed to terminate with no residual terms and thus the excess functional
has the desired property of giving the exact 0D free energy for any maximal cavity.

\subsection{One dimension}

	\begin{figure}
	              \centerline{\includegraphics[width=16cm]{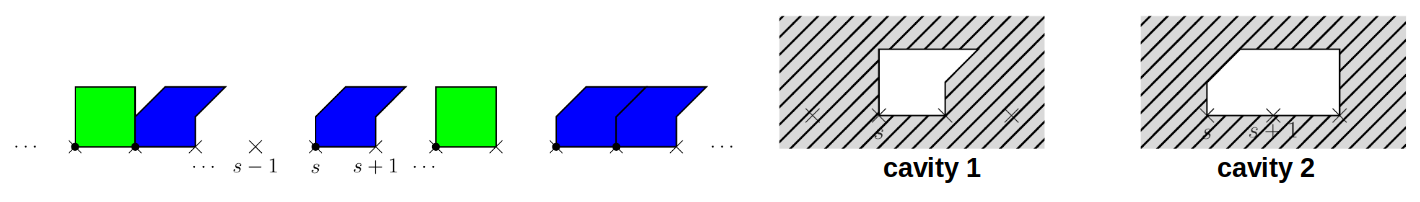}}
	  \caption{Visualization of the hard interactions in the mixture of colloids and polymer clusters in 1D. Maximal 0D cavities. }
	  \label{fig:1Ddef}
	\end{figure}

The 1D lattice gas had already been considered in Ref.~\cite{Cuesta2005}, nevertheless we include the derivation of its
excess free energy functional here for completeness.

In the 1D case there is just one species of polymers and polymer clusters beside the colloids
(see Fig.~\ref{fig:1Ddef}). 
With this representation the maximal zero dimensional cavities may be drawn as the hollow spaces in figure \ref{fig:1Ddef}.
These two maximal cavities imply to start off with the following weighted densities:
\begin{alignat}{2}
	n_1(s)&:=\rho_\text{pc}(s)+\rho_\text{c}(s)\quad&&=:\nOne{1}\notag\\
	n_2(s)&:=\rho_\text{pc}(s)+\rho_\text{c}(s+1)&&=:\nOne{2} \qquad.
	\label{eq:n1D}
\end{alignat}
Here, $\rhopc(s)=\sum_n \rho_{\text{pc}^n}(s)$ is the total density of polymer clusters and
we have introduced a graphical notation for the weighted densities, to make the fundamental measures easier to imagine. 
The first trial excess free energy functional then reads
\begin{equation}
	\beta \tFAO^\text{ex(1)}\left[ \rho_\text{c},\,\rho_\text{pc} \right] =
	\sums \left[ \pzd\left( \nOne{1}\right) +\pzd\left( \nOne{2}\right) \right] \qquad.
\end{equation}
Evaluating $\tFAO^\text{ex(1)}$ with the densities of the two  maximal cavities at an arbitrary location $r$ gives
\begin{alignat}{2}
	\beta \tFAO^\text{ex(1)}\left[ \cavOne{1}\right]&=\pzd\left( \nOne[r]{1}\right) && + \pzd\left( \nOne[r]{4}\right) + \pzd\left( \nOne[r]{3}\right)\notag\\
	\beta \tFAO^\text{ex(1)}\left[ \cavOne{2}\right]&= \pzd\left( \nOne[r]{2}\right) && + \pzd\left( \nOne[r]{3}\right)+ \pzd\left( \nOne[s]{4}\right)\qquad,
\end{alignat}
where we have naturally supplemented our graphical notation by the densities $\rho_\text{c}(s)=\nOne{4}$ and ${\rho_\text{pc}(s)=\nOne{3}}$. 
The first terms are the ones we want and we need to cancel the following two residual terms. 
In both cases this is achieved by adding $-\pzd\left( \nOne[s]{3}\right) - \pzd\left( \nOne[s]{4}\right)$ to our first trial free energy density. 
Then the improved second trial is given by:
\begin{equation}
	\beta \tFAO^\text{ex(2)}\left[ \rho_\text{c},\,\rho_\text{pc} \right] =
	\sums \left[ \pzd\left( \nOne{1}\right) +\pzd\left( \nOne{2}\right) -\pzd\left( \nOne[s]{3}\right) -  \pzd\left( \nOne[s]{4}\right)\right] \qquad.
\end{equation}
This produces no further spurious contributions when evaluated for the cavities of Fig.~\ref{fig:1Ddef} and consequently is our final excess free energy 
functional. 
By further defining the weighted densities
\begin{equation}
	n_3(s):=\rho_\text{pc}(s), \qquad n_4(s):=\rho_\text{c}(s)\qquad,
\end{equation}
the excess free energy density in $\tFAO^\text{ex}=\sum_s \Phi_\text{AO,1D}(s)$ reads
\begin{equation}
	\Phi_\text{AO,1D}(s)=\pzd\left( n_1\right) +\pzd\left( n_2\right) -\pzd\left( n_3\right) - \pzd\left( n_{4}\right)\qquad. \label{eq:Phi1D}
\end{equation}

	\begin{figure}
		\centerline{\includegraphics[width=12cm]{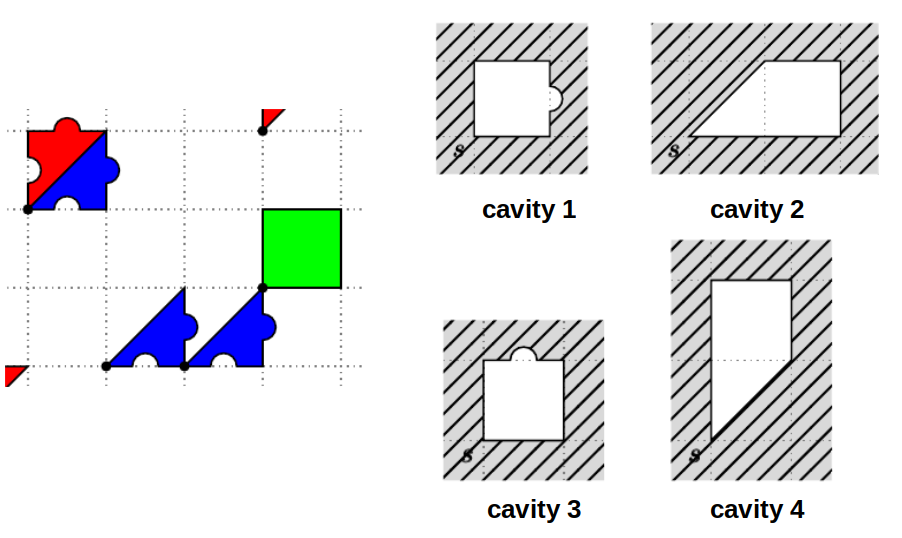}}
	  \caption{Visualization of the hard interactions in the mixture of colloids and polymer clusters in 2D and 
		maximal 0D cavities. }
	  \label{fig:2Ddef}
	\end{figure}

\subsection{Two dimensions}

In the 2D case there are two species of polymer clusters and colloids, their hard interactions are
visualized in Fig.~\ref{fig:2Ddef} (left part). There are four maximal cavities, these are shown in Fig.~\ref{fig:2Ddef} (right part).
These motivate to start the construction with the following weighted densities:
\begin{alignat}{2}
	n_1(\pmb{s})&:= \rho_{\text{pc},x}(\pmb{s})+\rho_\text{c}(\pmb{s})&&=:\nTwo{1} \notag \\
	n_2(\pmb{s})&:= \rho_{\text{pc},x}(\pmb{s})+\rho_\text{c}(\pmb{s}+\hat{\pmb{e}}_x)&&=:\nTwo{2} \notag \\
	n_3(\pmb{s})&:= \rho_{\text{pc},y}(\pmb{s})+\rho_\text{c}(\pmb{s})&&=:\nTwo{3} \notag \\
	n_4(\pmb{s})&:= \rho_{\text{pc},y}(\pmb{s})+\rho_\text{c}(\pmb{s}+\hat{\pmb{e}}_y)&&=:\nTwo{4} \qquad .
    \label{eq:n2D}
\end{alignat}
As in the one--dimensional case, we have defined an obvious graphical notation for the weighted densities. {Again we introduced a total polymer cluster density $\rho_{\text{pc},\alpha}=\sum_n\rho_{\text{pc}^n,\alpha}$.}
Our first ansatz for the free energy functional therefore is as follows:
\begin{equation}
	\beta \tFAO^\text{ex(1)} \left[ \rho_\text{c},\,\{\rho_{\text{pc},\alpha} \}\right] =
	\sum_{\pmb{s} \in \Lambda}\left[ \pzd\left( \nTwo{1}\right) +\pzd\left( \nTwo{2}\right) +\pzd\left( \nTwo{3}\right)+\pzd\left( \nTwo{4}\right)\right] \qquad.
\end{equation}
This trial functional is tested on the cavities in figure \ref{fig:2Ddef}. We obtain
\begin{alignat}{2}
\beta \tFAO^\text{ex(1)}\left[ \cavTwo{1}\right]&=\pzd\left( \nTwo[\pmb{r}]{1}\right) &&+ \pzd\left( \nTwo[\pmb{r}]{5}\right) +3\, \pzd\left( \nTwo[\pmb{r}]{7}\right) \notag\\
\beta \tFAO^\text{ex(1)}\left[ \cavTwo{2}\right]&= \pzd\left( \nTwo[\pmb{r}]{2}\right) &&+ \pzd\left( \nTwo[\pmb{r}]{5}\right) + 3\,\pzd\left( \nTwo[\pmb{s}]{7}\right)\notag\\
\beta \tFAO^\text{ex(1)}\left[ \cavTwo{3}\right]&= \pzd\left( \nTwo[\pmb{r}]{3}\right) &&+ \pzd\left( \nTwo[\pmb{r}]{6}\right) + 3\,\pzd\left( \nTwo[\pmb{r}]{7}\right)\notag\\
\beta \tFAO^\text{ex(1)}\left[ \cavTwo{4}\right]&= \pzd\left( \nTwo[\pmb{r}]{4}\right) &&+ \pzd\left( \nTwo[\pmb{r}]{6}\right) + 3\,\pzd\left( \nTwo[\pmb{s}]{7}\right)\qquad,
\end{alignat}
with the definitions $\rho_\text{c}=:\nTwo[ ]{7}$, $\rho_{\text{pc},x}=:\nTwo[ ]{5}$ and $\rho_{\text{pc},y}=:\nTwo[ ]{6}$. 
For each cavity, the first term is the desired contributionand the following two are spurious. 
As before, we eliminate them by subtracting 
$\sum\nolimits_{\pmb{s}}\left[ \pzd\left( \nTwo{5}\right) +\pzd\left( \nTwo{6}\right) +3\,\pzd\left( \nTwo{7}\right)\right]$ 
from the free energy functional $\beta \tFAO^\text{ex(1)}$. The resulting functional
\begin{multline}
\beta \tFAO^\text{ex(2)} \left[ \rho_\text{c},\,\{\rho_{\text{pc}i} \}\right] =
\sum_{\pmb{s} \in \Lambda}\left[ \pzd\left( \nTwo{1}\right) +\pzd\left( \nTwo{2}\right) +\pzd\left( \nTwo{3}\right)+\pzd\left( \nTwo{4}\right)\right. \\
-\left. \pzd\left( \nTwo{5}\right) -\pzd\left( \nTwo{6}\right) -3\,\pzd\left( \nTwo{7}\right)\right] \qquad.
\end{multline}
is the final functional, since no further spurious contributions are produced when evaluated for the maximal cavities.
Defining the weighted densities
\begin{equation}
	n_5(\pmb{s}):=\rho_{\text{pc},x}(\pmb{s}), \qquad n_6(\pmb{s}):=\rho_{\text{pc},y}(\pmb{s}), \qquad
	n_7(\pmb{s}):=\rho_{\text{c}}(\pmb{s})\qquad,
\end{equation}
the excess free energy density in $\beta \tFAO^\text{ex}=\sum_{\pmb{s}} \Phi_\text{AO,2D}(\pmb{s})$ is given by
\begin{multline}
	\Phi_\text{AO,2D}(\pmb{s})=\pzd\left( n_1\right) +\pzd\left( n_2\right) +\pzd\left( n_3\right) + \pzd\left( n_{4}\right) \\
	- \pzd\left( n_{5}\right) - \pzd\left( n_{6}\right) - 3\,\pzd\left( n_{7}\right)\qquad. 
	\label{eq:Phi2Da}
\end{multline}

	\begin{figure}
		\centerline{\includegraphics[width=12cm]{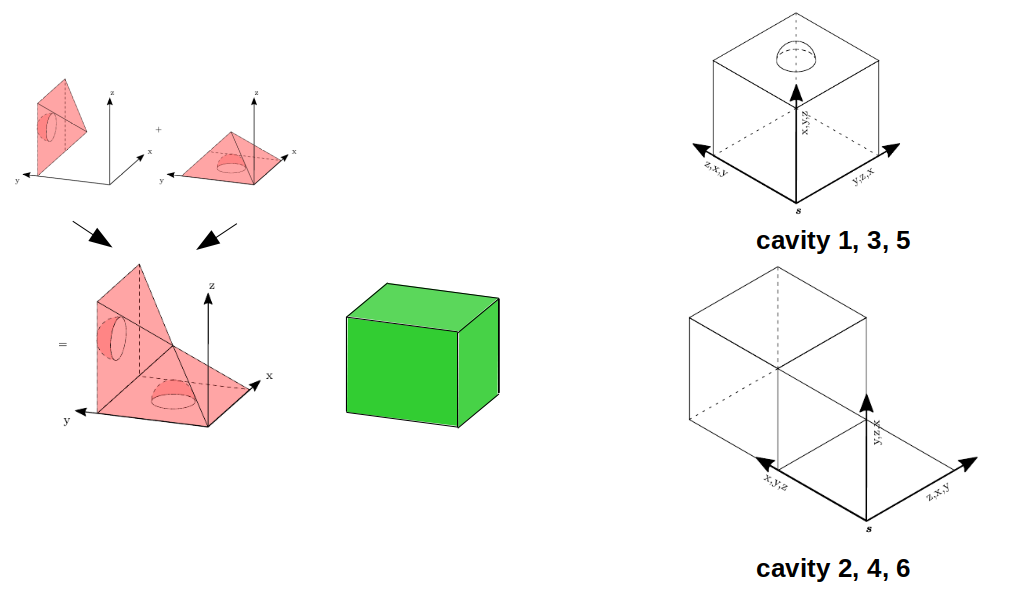}}
		\caption{Visualization of the hard interactions in the mixture of colloids (green cube) and 
		polymer clusters (of type pc$,y$) in 3D.
		The polymer cluster pc$,y$ consist of two pyramids (upper left part) with a square base glued together 
		on their side surfaces. One of the pyramids has a hemispherical cutout in the center of its base and the other a 
		hemispherical bulge (as shown). The bulge forbids a colloid (green cube) to sit next to the cluster 
		in the direction of the bulge. The other two polymer clusters pc$,x$ and pc$,z$ look same in shape
		but differ in their orientation. (The right orientation can be obatined by permutig the coordinate
		axes: $(xyz) \to (zxy)$ for pc$,x$ and $(xyz) \to (yzx)$ for pc$,z$.) 
		For polymer clusters of different species, the hemisphere fits exactly into the cutout so that the 
		clusters can sit side by side, and the pyramid geometry of the clusters enables them to simultaneously sit 
		on the same lattice site. 
		Maximal zero-dimensional cavities (right): The cavities 1, 3 and 5 differ only in the direction the bulge points. 
		They can either host a colloid or a certain polymer cluster at the lattice site $\pmb{s}$. 
		The cavities 2, 4 and 6 of figure (b) also differ only in their orientation and can either host a certain polymer 
		cluster at the lattice site $\pmb{s}$ or a colloid at the lattice site $\pmb{s}+\hat{\pmb{e}}_{x/y/z}$. }
	  \label{fig:3Ddef}
	\end{figure}

\subsection{Three dimensions}

In the 3D case there are three species of polymer clusters and colloids, their hard interac-
tions are visualized in Fig.~\ref{fig:3Ddef} (left part). There are six maximal cavities, these are shown
in a reduced form in Fig.~\ref{fig:3Ddef} (right part). These motivate to start the construction with the following weighted
densities:

\begin{align}
	m_1(\pmb{s})&:=\rho_{\text{pc},x}(\pmb{s})+\rho_\text{c}(\pmb{s}) \notag\\
	m_2(\pmb{s})&:=\rho_{\text{pc},x}(\pmb{s})+\rho_\text{c}(\pmb{s}+\hat{\pmb{e}}_x) \notag\\
	m_3(\pmb{s})&:=\rho_{\text{pc},y}(\pmb{s})+\rho_\text{c}(\pmb{s}) \notag\\
	m_4(\pmb{s})&:=\rho_{\text{pc},y}(\pmb{s})+\rho_\text{c}(\pmb{s}+\hat{\pmb{e}}_y) \notag\\
	m_5(\pmb{s})&:=\rho_{\text{pc},z}(\pmb{s})+\rho_\text{c}(\pmb{s}) \notag\\
	m_6(\pmb{s})&:=\rho_{\text{pc},z}(\pmb{s})+\rho_\text{c}(\pmb{s}+\hat{\pmb{e}}_z) \qquad.
    \label{eq:n3D}
\end{align}
A graphical notation of the weighted densities is not particulary helpful. The first ansatz for the free energy functional thus reads
\begin{multline}
	\beta \tFAO^\text{ex(1)}\left[ \rho_\text{c},\,\{\rho_{\text{pc},\alpha} \}\right] =
	\sum_{\pmb{s} \in \Lambda}\left[ \Phi^{(0)}\left( m_1(\pmb{s})\right) + \Phi^{(0)}\left( m_2(\pmb{s})\right) + \Phi^{(0)}\left( m_3(\pmb{s})\right) \right. \\
	\left. + \Phi^{(0)}\left( m_4(\pmb{s})\right) + \Phi^{(0)}\left( m_5(\pmb{s})\right) + \Phi^{(0)}\left( m_6(\pmb{s})\right) \right] \qquad.
\end{multline}
Evaluating for the density profiles $\rho_{\text{cav},i} \; (i=1,\hdots, 6$) of the 0D cavities of Fig.~\ref{fig:3Ddef}, 
one obtains
\begin{alignat}{2}
\beta \tFAO^\text{ex(1)} \left[ \rho_{\text{cav},1}\right]&=\pzd\left( m_1(\pmb{s})\right) &&+ \pzd\left( \rho_{\text{pc}, x}(\pmb{s})\right) +5\, \pzd\left( \rho_\text{c}(\pmb{s})\right) \notag\\
\beta \tFAO^\text{ex(1)}\left[ \rho_{\text{cav},2}\right]&=\pzd\left( m_2(\pmb{s})\right) &&+ \pzd\left( \rho_{\text{pc}, x}(\pmb{s})\right) +5\, \pzd\left( \rho_\text{c}(\pmb{s}+\hat{\pmb{e}}_x)\right) \notag\\
\beta \tFAO^\text{ex(1)}\left[ \rho_{\text{cav},3}\right]&=\pzd\left( m_3(\pmb{s})\right) &&+ \pzd\left( \rho_{\text{pc}, y}(\pmb{s})\right) +5\, \pzd\left( \rho_\text{c}(\pmb{s})\right) \notag\\
\beta \tFAO^\text{ex(1)}\left[ \rho_{\text{cav},4}\right]&=\pzd\left( m_4(\pmb{s})\right) &&+ \pzd\left( \rho_{\text{pc}, y}(\pmb{s})\right) +5\, \pzd\left( \rho_\text{c}(\pmb{s}+\hat{\pmb{e}}_y)\right) \notag\\
\beta \tFAO^\text{ex(1)}\left[ \rho_{\text{cav},5}\right]&=\pzd\left( m_5(\pmb{s})\right) &&+ \pzd\left( \rho_{\text{pc}, z}(\pmb{s})\right) +5\, \pzd\left( \rho_\text{c}(\pmb{s})\right) \notag\\
\beta \tFAO^\text{ex(1)}\left[ \rho_{\text{cav},6}\right]&=\pzd\left( m_6(\pmb{s})\right) &&+ \pzd\left( \rho_{\text{pc}, z}(\pmb{s})\right) +5\, \pzd\left( \rho_\text{c}(\pmb{s}+\hat{\pmb{e}}_z)\right)\qquad.
\end{alignat}
Also here spurious contributions in addition to the desired contribution (first term on the rhs, respectively) arise. 
An improved functional follows by subtracting these spurious contributions:
\begin{align}
	\beta \tFAO^\text{ex(2)}\left[ \rho_\text{c},\,\{\rho_{\text{pc},\alpha} \}\right] =
	\sum_{\pmb{s} \in \Lambda} \Bigl[ &\pzd\left( m_1(\pmb{s})\right) + \pzd\left( m_2(\pmb{s})\right) + \pzd\left( m_3(\pmb{s})\right) + \pzd\left( m_4(\pmb{s})\right) \notag\\
	+ &\pzd\left( m_5(\pmb{s})\right) + \pzd\left( m_6(\pmb{s})\right) - \pzd\left( \rho_{\text{pc}, x}(\pmb{s})\right)\notag\\
	- &\pzd\left( \rho_{\text{pc}, y}(\pmb{s})\right) - \pzd\left( \rho_{\text{pc}, z}(\pmb{s})\right)-5\, \pzd\left( \rho_\text{c}(\pmb{s})\right)\Bigr] \qquad.
\end{align}
This functional is the final result since the evaluation for the 0D cavity densities $\rho_{\text{cav},i}$
gives the exact result $\pzd\left( m_i\right)$ .
We define the additional weighted densities
\begin{alignat}{2}
	m_7(\pmb{s}):=&\rho_{\text{pc},x}(\pmb{s})\qquad \qquad m_8(\pmb{s})&&:=\rho_{\text{pc},y}(\pmb{s}) \notag \\
	m_9(\pmb{s}):=&\rho_{\text{pc},z}(\pmb{s})\qquad \qquad m_{10}(\pmb{s})&&:=\rho_{\text{c}}(\pmb{s})\qquad.
\end{alignat}
The excess free energy density in $\beta \tFAO^\text{ex}=\sum_{\pmb{s}} \Phi_\text{AO,3D}(\pmb{s})$ is given by
\begin{multline}
 \Phi_\text{AO,3D}(\pmb{s})=\pzd\left( m_1\right) +\pzd\left( m_2\right) +\pzd\left( m_3\right) + \pzd\left( m_{4}\right) + \pzd\left( m_{5}\right)+ \pzd\left( m_{6}\right)\\
- \pzd\left( m_{7}\right) - \pzd\left( m_{8}\right) - \pzd\left( m_{9}\right)- 5\,\pzd\left( m_{10}\right)
	\label{eq:Phi3Da} \;.
\end{multline}

%% file: app_C_surfacetension.tex
\section{Surface tensions from DFT}
\label{app:C}

In general, a surface or interface tension is defined in DFT as an excess grand potential
per unit area $\gamma = (\Omega - \Omega_\text{bulk})/A$ where $\Omega_\text{bulk}$ is a
bulk grand potential. For a liquid--vapour interface, it is the grand potential of either coexisting
state and for a fluid next to a wall, it is the grand potential of the asymptotic bulk state far away from
the wall.
\begin{figure}
    \centering
    \includegraphics[width=0.45\textwidth]{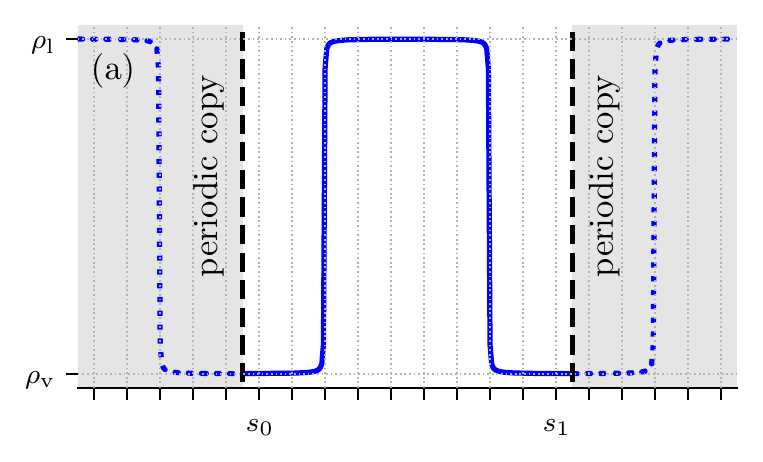}
    \includegraphics[width=0.45\textwidth]{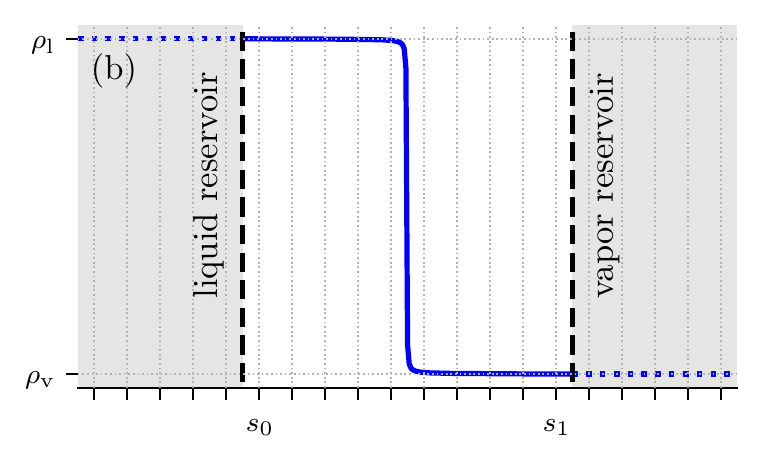}
    \includegraphics[width=0.45\textwidth]{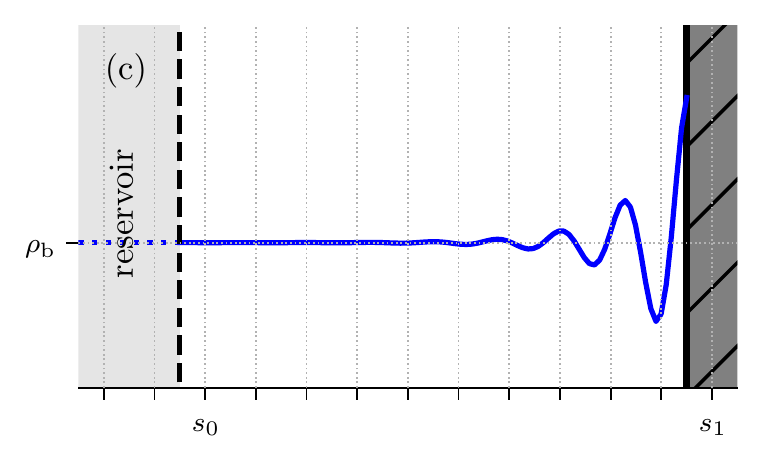}
    \includegraphics[width=0.45\textwidth]{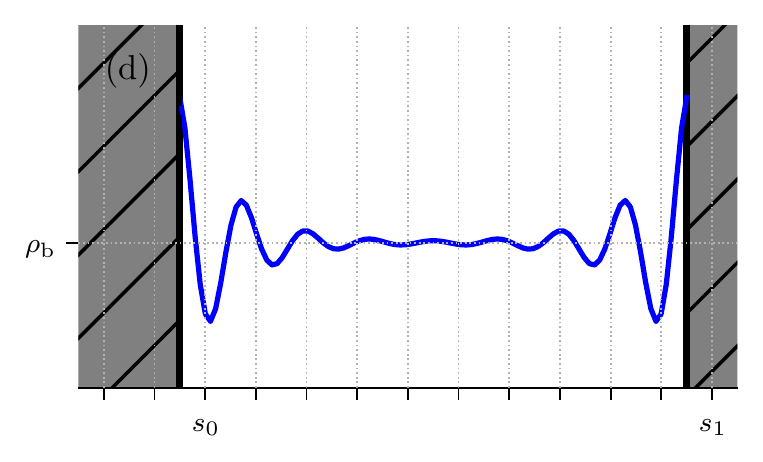}
    \caption{Sketch of different boundary conditions. In figure (a) periodic boundary conditions are chosen. In figure (b) the computational box is connected to a liquid reservoir at the lower end of the box and a vapor reservoir at the upper end. Figure (c) shows a geometry where the system is connected to a bulk reservoir on the left and an infinite potential (wall) constrains the system to the right. In figure (d) two walls constrain the computational box.}
    \label{fig:sketch_bc}
\end{figure}
Using a box with periodic boundary conditions (and consequently two interfaces, see Fig.~\ref{fig:sketch_bc}(a)), there is no
problem in determining the excess grand potential per unit area as it stands and it
corresponds to twice the surface tension.
For a generic, nonsymmetric situation as depicted in Fig.~\ref{fig:sketch_bc}(b), a peculiar problem arises when
the surface tension is evaluated in a finite computational box which to our knowledge has not been described
before and which is connected to the use of nonsymmetric weight functions in density functionals.
This problem is also apparent when one looks at the grand potential density in the periodic situation: it
is not symmetric for the two interfaces and therefore it is not clear how to evaluate the
surface tension for one interface only.

In order to address that problem,
we assume generically an FMT form for the free energy in which the free energy density is composed
of terms local in densities (i.e. ideal gas and external potential contributions) and other terms
local in weighted densites but nonlocal in densities
\bea
  \beta \F [\rho_i(\pmb{s})] = \sums \left( \Phil(\rho_i(\pmb{s})) + \Phinl[\rho_i(\pmb{s})] \right) \;.
\eea
In our specific case, $\rho_i = \{\rhoc,\rhopca\}$, and there are two types of nonlocal terms. 
The first type is in the excess free energy of colloids and polymer clusters and has the form
$\Phinl_{(1),\alpha}(n^{(\alpha)}(\pmb{s}))=\pzd(n^{(\alpha)}(\pmb{s}))$ with $n^{(\alpha)}(\pmb{s}) = \rhopca(\pmb{s}) + 
\rhoc(\pmb{s}+\hat{\pmb{e}}_\alpha)$ (see Eqs.~(\ref{eq:Phi1D},\ref{eq:Phi2Da},\ref{eq:Phi3Da}) for the
free energy densities and Eqs.~(\ref{eq:n1D},\ref{eq:n2D},\ref{eq:n3D}) for the weighted densities).
The second type comes from the subtraction of the colloid one--body term (see Eq.~(\ref{eq:flg_final}))
and has the form 
$ \Phinl_{(2),\alpha} (\pmb{s})=- \rhopr \;\rhoc(\pmb{s}+\hat{\pmb{e}}_\alpha)$.
We need to consider only those terms where  the direction $\alpha$ is perpendicular to the interface
or wall.

We consider explicitly the situation with the fluid next to a wall (see Fig.~\ref{fig:sketch_bc}(c) for the asymmetric situation and the
corresponding symmetrized one in Fig.~\ref{fig:sketch_bc}(d)) whose resolution also solves
the problem for the liquid-vapor interface. Let ${s}_{0[1]}$ be the lower[upper] limit of the one--dimensional computational box.
The upper limit is assumed to be deep in the wall where all densities and free energy densities are zero.
A naive evaluation of the surface tension is given by
\bea
  \gamma(\mu_i;[\rho_j(\mu_i)]) &=& \sum_{{s}={s}_0}^{{s}_1} \left( \Phil(\rho_i({s})) - \sum_i \mu_i \rho_i({s}))  - \omega_\text{b} (\mu_i) \right)
  + \sum_{{s}={s}_0}^{{s}_1} \Phinl[\rho_i({s})]
\eea
where $\mu_i$ is the bulk chemical potential of species $i$ and $\omega_\text{b}(\mu_i)=-p(\mu_i)$ is the grand
potential density of the asymptotic bulk state. Owing to the asymmetry of the grand potential density mentioned above,
this gives a wrong result for $\gamma$. The failure becomes manifest if we consider the Gibbs adsorption
equation
\bea
  \frac{d\gamma}{d\mu_i} = -\Gamma_i = - \sum_{{s}={s}_0}^{{s}_1} (\rho_i({s})
  - \rho_{i,\text{b}})
\eea
{(with $\rho_{i,\text{b}}$ being the bulk density of species $i$)} which should be  fulfilled in DFT as can be seen as follows:
\bea
  \frac{d\gamma}{d\mu_i} &=& \frac{\partial \gamma}{\partial \mu_i} +
  \sum_s \sum_j \frac{\partial \gamma}{\partial \rho_j({s}) } \frac{\partial \rho_j({s}) }{\partial \mu_i} \\
  &=& -\sum_s \rho_i({s}) + \sum_s \rho_{i,\text{b}}   
\eea
(sums over ${s}$ are unrestricted). Here we have used the Gibbs--Duhem relation 
$\partial p/\partial \mu_i=\rho_{i,\text{b}}$ and the grand potential minimization condition
$\partial \Omega/\partial \rho_j({s}) = 0 = \partial \gamma/\partial \rho_j({s})  $.

However, in the finite box the derivative $\partial \gamma/\partial \rho_j({s})$ is not zero for
points close to the lower boundary. In our specific case it affects only the derivative with respect to
$\rhoc(s_0)$ which should pick up a term from the grand potential density at point $s_0-1$ (which is absent). Thus
we find
\bea
   \frac{\partial (\beta \gamma)}{\partial \rhoc({s_0})} = - \frac{\partial \pzd(\rhocb+\rhopcb)}{\partial \rhocb}
   + \rhopr\;. 
\eea
Here we have assumed that at the lower end of the bulk, all densities are at their bulk values. This can be remedied
by adding a term $X$ to the surface tension, $\beta \gamma \to \beta \gamma+X$, with the following properties:
\bea
  \frac{\partial X}{\partial \rhocb} &=&  \frac{\partial \pzd(\rhocb+\rhopcb)}{\partial \rhocb} - {\rhopr}\quad \text{and} \\
   \frac{\partial X}{\partial \rhopcb} &=& 0 \;.
\eea
The term $X$ can be constructed from the bulk free energy density $f$ which has the properties
$\partial f/\partial \rhoc = \mu_\text{c}$ and 
$\partial f/\partial \rhopca = \partial f/\partial \rhopcb   = \beta^{-1} \ln\zeta$.  
It is given by
\bea
 X = \frac{1}{2} \Phi^\text{id} (\rhopcb) + \pzd(\rhocb+\rhopcb) - \frac{1}{2} \pzd(\rhopc)
    - \frac{1}{2} \rhopcb  \ln\zeta  - \rhopr \;\rhocb\;, 
\eea
which is valid in all three dimensions. For the problem of the single
liquid--vapour interface, a second term has to be subtracted (coming from the upper limit of the box)
which has the same form as $X$ but differs in the bulk densities (which have to be the ones at the upper limit).
Via this route, $X$ is only determined up to a constant but one
can check the new surface tension numerically by comparing to the results in the box with periodic boundary 
condition which show that the constant is zero.

The problem with the surface tension for a single surface appears generically in DFT models with nonsymmetric weight
functions, such as e.g. in the lattice rod functionals of Refs.~\cite{Oettel2016,Mortazavifar2017}.
For symmetric weight functions one can check that missing terms in the derivative $\partial \gamma/\partial \rho_j({s})$ near the boundaries are cancelled by additional terms coming from derivatives with respect to
$\rho_j({s})$ just outside the box limits. The usual continuum FMT functionals work with symmetric weight functions
such that this problem never arose in previous computations.

%% file: ms.bbl
\begin{thebibliography}{10}

\bibitem{Huang1987}
K.~Huang,
\newblock {\em Statistical Mechanics}, 2 ed. (John Wiley \& Sons, 1987).

\bibitem{Landau2018}
A.~M. Ferrenberg, J.~Xu, and D.~P. Landau,
\newblock Phys. Rev. E {\bf 97}, 043301 (2018).

\bibitem{Nishino1999}
T.~Nishino, K.~Okunishi, Y.~Hieida, T.~Hikihara, and H.~Takasaki,
\newblock Transfer-matrix approach to classical systems,
\newblock in {\em Density--Matrix Renormalization (Lecture Notes in Physics
  vol. 528)}, edited by I.~Peschel, M.~Kaulke, X.~Wang, and K.~Hallberg,
  chap.~5, p. 127, Springer, Berlin, Heidelberg, 1999.

\bibitem{Drzewinski2009}
A.~Drzewi{\' n}ski, A.~Macio{\l}ek, A.~Barasi{\' n}ski, and S.~Dietrich,
\newblock Phys. Rev. E {\bf 79}, 041145 (2009).

\bibitem{Drzewinski2012}
M.~Zubaszewska, A.~Gendiar, and A.~Drzewi{\' n}ski,
\newblock Phys. Rev. E {\bf 86}, 062104 (2012).

\bibitem{Asakura1954}
S.~Asakura and F.~Oosawa,
\newblock J. Chem. Phys. {\bf 22}, 1255 (1954).

\bibitem{Asakura1958}
S.~Asakura and F.~Oosawa,
\newblock J. Polymer Science {\bf 33}, 183 (1958).

\bibitem{Cuesta2005}
J.~A. Cuesta, L.~Lafuente, and M.~Schmidt,
\newblock Phys. Rev. E {\bf 72}, 031405 (2005).

\bibitem{Lafuente2002}
L.~Lafuente and J.~A. Cuesta,
\newblock J. Phys.: Condensed Matter {\bf 14}, 12079 (2002).

\bibitem{Lafuente2004}
L.~Lafuente and J.~A. Cuesta,
\newblock Phys. Rev. Lett. {\bf 93}, 130603 (2004).

\bibitem{Archer2014}
A.~P. Hughes, U.~Thiele, and A.~J. Archer,
\newblock Am. J. Phys. {\bf 82}, 1119 (2014).

\bibitem{Troester2005}
A.~Tr\"oster and C.~Dellago,
\newblock Phys. Rev. E {\bf 71}, 066705 (2005).

\bibitem{Troester2005a}
A.~Tr\"oster, C.~Dellago, and W.~Schranz,
\newblock Phys. Rev. B {\bf 72}, 094103 (2005).

\bibitem{Evans1979}
R.~Evans,
\newblock Adv. Phys. {\bf 28}, 143 (1979).

\bibitem{Bragg1934}
W.~L. Bragg and E.~J. Williams,
\newblock Proc. Roy. Soc. London A {\bf 145}, 699 (1934).

\bibitem{Onsager1944}
L.~Onsager,
\newblock Physical Review {\bf 65}, 117 (1944).

\bibitem{Mortazavifar2017}
M.~Mortazavifar and M.~Oettel,
\newblock Phys. Rev. E {\bf 96}, 032608 (2017).

\bibitem{Brader2003}
J.~M. Brader, R.~Evans, and M.~Schmidt,
\newblock Molecular Physics {\bf 101}, 3349 (2003).

\bibitem{Gschwind2017}
A.~Gschwind, M.~Klopotek, Y.~Ai, and M.~Oettel,
\newblock Phys. Rev. E {\bf 96}, 012104 (2017).

\bibitem{Bethe1935}
H.~A. Bethe,
\newblock Proc. Roy. Soc. London A {\bf 150}, 552 (1935).

\bibitem{Pan1995}
J.~Pan and S.~D. Gupta,
\newblock Phys. Rev. C {\bf 51}, 1384 (1995).

\bibitem{Berg1993}
B.~A. Berg, U.~Hansmann, and T.~Neuhaus,
\newblock Z. Phys. B {\bf 90}, 229 (1993).

\bibitem{Bittner2009}
E.~Bittner, A.~Nu{\ss}baumer, and W.~Janke,
\newblock Nuc. Phys. B {\bf 820}, 694 (2009).

\bibitem{dePablo2003}
T.~S. Jain and J.~J. de~Pablo,
\newblock J. Chem. Phys. {\bf 118}, 4226 (2003).

\bibitem{Troester2017}
A.~Tr\"oster, F.~Schmitz, P.~Virnau, and K.~Binder,
\newblock J. Phys. Chem. B {\bf 122}, 3407 (2017).

\bibitem{Lin2020}
S.-C. Lin, G.~Martius, and M.~Oettel,
\newblock J. Chem. Phys. {\bf 152}, 021102 (2020).

\bibitem{Lafuente2005}
L.~Lafuente and J.~Cuesta,
\newblock J. Phys. A: Math. Gen. {\bf 38}, 7461 (2005).

\bibitem{Oettel2016}
M.~Oettel, M.~Klopotek, M.~Dixit, E.~Empting, T.~Schilling, and H.~Hansen-Goos,
\newblock J. Chem. Phys. {\bf 145}, 074902 (2016).

\end{thebibliography}
